\documentstyle[12pt]{article}

\newcommand{\be}{\begin{equation}}
\newcommand{\ee}{\end{equation}}
\newcommand{\bea}{\begin{eqnarray}}
\newcommand{\eea}{\end{eqnarray}}

\newcommand{\ba}{\begin{array}}
\newcommand{\ea}{\end{array}}
\newcommand{\vs}[1]{\vspace{#1 mm}}

\def\bbox{{\,\lower0.9pt\vbox{\hrule \hbox{\vrule height 0.2 cm
\hskip 0.2 cm \vrule height 0.2 cm}\hrule}\,}}
\newcommand{\dsl}{\pa \kern-0.5em /}

\newcommand{\pa}{\partial}

\font\mybb=msbm10 at 10pt
\def\bb#1{\hbox{\mybb#1}}

\def\bE {\bb{E}}

\begin{document}

\topmargin 0pt
\oddsidemargin 0mm

\renewcommand{\thefootnote}{\fnsymbol{footnote}}
\begin{titlepage}

\begin{flushright}
DAMTP-1998-137\\
hep-th/9810041\\
\end{flushright}
\vs{15}

\begin{center}
{\Large\bf On Gauged Maximal Supergravity In Six
Dimensions\footnote{e-mail address: p.m.cowdall@damtp.cam.ac.uk}}
\vs{10}

{\large P.M. Cowdall} \\
\vs{5}
{\em DAMTP, University of Cambridge,\\ 
Silver Street, Cambridge CB3 9EW, UK.}
\end{center}
\vs{10}
\centerline{{\bf Abstract}}

The maximal SO(5) gauged D=7 supergravity is
dimensionally reduced to six dimensions yielding a new SO(5) gauged
D=6 model. It is shown that, unlike in D=7, the SO(5) gauge coupling
constant can be taken to zero to yield the maximally extended
supergravity in six dimensions. It is also shown
that the limit of D=5 N=4 SU(2)$\times$U(1) gauged supergravity in
which the U(1) coupling constant is turned off can be obtained.

\end{titlepage}
\newpage 
\renewcommand{\thefootnote}{\arabic{footnote}}
\setcounter{footnote}{0}

\section{Introduction}

Due to the recently conjectured AdS/CFT correspondence \cite{malda} and its
generalisation, the domain-wall/QFT correspondence \cite{BST}, the
determination of all possible gaugings of higher dimensional
supergravities is currently attracting much interest. Gauged maximal
supergravities are known to exist in dimensions 4,$\,$5,$\,$7 and
8,$\,$ see refs \cite{DWN}--\cite{SS1}. 
There is a clear lack of a gauged maximal
supergravity in six dimensions, thus the presentation of such a model 
will be the focus of this paper.

To understand the construction of a gauged maximal supergravity in D=6
we begin with the ungauged model of Tanii \cite{Tanii} for which the bosonic
field content is 1 graviton, 5 second rank antisymmetric tensor
potentials, 16 vectors and 25 scalars. The model has (4,4) maximal
supersymmetry {\it i.e.} 4 spinor charges of each chirality, and the
automorphism group\footnote{Usp(4)$\simeq$Spin(5), the covering group
of SO(5).} of the superalgebra is SO(5)$\times$SO(5). The scalars
parametrise the coset space SO(5,5)/SO(5)$\times$SO(5). The 16 vector
fields are in the $\underline{16}$ spinor representation of SO(5,5)
which decomposes into the (4,4) representation of SO(5)$\times$SO(5).
Clearly there are not enough vectors to gauge SO(5)$\times$SO(5). 
However, the (4,4) representation of SO(5)$\times$SO(5) decomposes w.r.t. its
diagonal SO(5) subgroup in the following way
\be
(4,4) = \underline{10}+\underline{5}+\underline{1}.
\ee
The $\underline{10}$ could be used to gauge SO(5), but there are
potential consistency problems with the $\underline{5}\,$; one may try
to solve these problems using constructions similar to those in D=7
and D=5 \cite{PKT},
but these only work in odd dimensions. Alternatively one could try to
gauge the larger group ISO(5) for which the
$\underline{10}+\underline{5}$ of SO(5) is the adjoint
$\underline{15}$. There also remains the question of what the ground
state of this D=6 gauged model will be; due to the lack of a maximally
extended supergroup in six dimensions, there cannot be a maximally
supersymmetric AdS$_6$ vacuum.

The solution to the gauging of Tanii's model is provided by the
S$^1$ reduction of the D=7 SO(5) gauged supergravity \cite{PPvN}. 
The bosonic field content is 1
graviton, 5 third rank massive self dual antisymmetric tensor
potentials, 10 vectors and 14 scalars. 
One feature of D=7 SO(5) gauged supergravity that is relevant is that
the scalar potential has a supersymmetry preserving maximum, leading
to an AdS$_7$ vacuum with isometry supergroup OSp(6,2$\mid$4)
\cite{PPvNW}. Thus there is no M$_6{\times}$S$^1$ vacuum solution and so
dimensional reduction of this model was thought to be inconsistent. In
fact as was pointed out in \cite{LPT} the reduction is perfectly consistent
and is implemented using the standard Kaluza-Klein (KK) ansatz on the
fields; substitution of the standard KK ansatz for the fields into the
lagrangian gives a lower dimensional lagrangian whose field equations
are the same as those obtained by direct substitution of the KK ansatz
into the higher dimensional field equations. Having said that the
reduction is always consistent, one is not guaranteed solutions in the
lower dimension unless there exists a solution with a U(1) isometry in
the higher dimension. In the case of D=7 maximal SO(5) gauged
supergravity, the AdS$_7$ ground state has a U(1) isometry and, as
shown in \cite{LPT}, upon reduction becomes a D=6 domain wall preserving 1/2
of the supersymmetry. Hence, there is no
obstacle to the S$^1$ reduction of D=7 SO(5) gauged supergravity. 

Proceeding with the reduction it is clear that the gauge group in D=6 
will also be
SO(5). The 10-plet of vectors survives in D=6 and is supplemented by a
single KK vector. Hence there are 11 vectors in D=6 transforming under
SO(5) as a $\underline{10}+\underline{1}$. The 5-plet of third rank
massive self dual antisymmetric tensor potentials splits into a 5-plet
of massive 3-index potentials and a 5-plet of massive 2-index
potentials. As the self duality is lost in the reduction, the 3-index
potentials are auxiliary and can be eliminated. Hence in D=6 the
massive 2-index tensor potentials transform as a $\underline{5}\,$;
this accounts for the remaining 5 vectors and massless 2-index
potentials of the ungauged theory. Thus, understanding the SO(5)
gauged D=6 model is much easier from the D=7 perspective, as is its
construction.

Another interesting feature of D=7 SO(5) gauged supergravity is that
the global limit, in which the SO(5) gauge coupling constant is turned
off, cannot be obtained. One might suspect that this will also be true
in D=6. We find in fact that this is not the case and the global limit
can be taken smoothly allowing us to make contact with the maximal D=6
supergravity \cite{Tanii}.

The paper is set out as follows. In section two we perform the
dimensional reduction of the bosonic sector of D=7 SO(5) gauged
supergravity to six dimensions and obtain a new D=6 SO(5) gauged
model. In section three we show the $g{\rightarrow}0$ limit is
attainable after appropriate rescalings. In section four we present
our conclusions. In appendices A and C we give the complete
Lagrangians and supersymmetry transformation laws of D=6 SO(5)
supergravity and D=6 maximal supergravity respectively. On the subject
of taking limits, using a similar method, we also show in appendix D
that the limit of D=5 N=4 SU(2)$\times$U(1) gauged supergravity in
which the U(1) coupling constant is turned off can be obtained.

\section{Reduction to D=6}

First we recall some features of the construction of maximal SO(5)
gauged D=7 supergravity. One of these is the necessity for `odd
dimensional self duality' \cite{PPvN,PKT}. In the ungauged maximal D=7
model \cite{SS2} there are five third rank antisymmetric tensor
potentials $A_{\alpha\beta\gamma I}$. These fields are expected to
transform in the vector representation of SO(5) upon gauging but
simply replacing the ordinary derivative by an SO(5) covariant
derivative in the Maxwell action of $A_{\alpha\beta\gamma I}$ would
break the antisymmetric tensor gauge invariance.  Thus there would not
be a matching of bosonic and fermionic degrees of freedom necessary
for supersymmetry. A way around this problem is provided by `odd
dimensional self duality'. In seven dimensions a massless 3-index
antisymmetric tensor has the same number of degrees of freedom as a
massive self dual 3-index antisymmetric tensor, namely ten. Therefore
the Maxwell action of the massless $A_{\alpha\beta\gamma I}$ fields
can be replaced by an action whose field equation reads \be mS_3 =
{\epsilon}^7 dS_3 \ee hence the name self duality in odd
dimensions. Iteration of this equation yields a massive Proca
equation. Therefore the 3-index fields have only ten propagating
degrees of freedom. However now one is in a position to effect the
gauging as the ordinary derivative can simply be replaced by an SO(5)
covariant derivative. The antisymmetric tensor gauge invariance no
longer exists but isn't needed as the potentials $S_{\alpha\beta\gamma
I}$ each propagate the correct number of degrees of freedom required
by supersymmetry and also transform as a 5-plet under the local SO(5)
symmetry. The parameter m turns out to be proportional to the SO(5)
coupling constant g.

A side-effect of `odd dimensional self duality' is that there is a
`gauge discontinuity'. {\it i.e.} g appears non analytically in the
Chern-Simons terms and in the supersymmetry transformation law of
$S_{\alpha\beta\gamma I}$; hence no $g{\rightarrow}0$ limit exists.
We will see that after reduction to D=6 the
$g{\rightarrow}0$ limit can be taken. We begin with the bosonic sector
of D=7 maximal SO(5) gauged supergravity, which is \cite{PPvN}\footnote{We note
that the sign of the mass term of the 3-index field is opposite to the
one in \cite{PPvN}. That this is the correct sign can be checked by choosing
S$_{123}=u$, S$_{456}=v$, where $u$ and $v$ depend only on time, and letting
all other components of S$_{\alpha\beta\gamma}$ vanish. With the
choice  ${\Pi}^{-1 \; I}_{\; \; \; \; j}= {\delta}^{I}_{j}$, the
lagrangian (neglecting interaction terms) for the 3-index field then
becomes ${{\dot v}\over{2}}^{2}-{{m}\over{2}}^{2}v^{2}$ after
elimination of the auxiliary field $u$. Hence has the required form
KE$-$PE.}

\bea
e^{-1}{\cal L}_7 &=& {R\over2} -{1\over4}({\Pi}_{I}^{\; \; i}{\Pi}_{J}^{\;
\; j}F_{\alpha\beta}^{IJ})^2 -{{m^2}\over2}({\Pi}^{-1 \; I}_{\; \;
\; \; i}S_{\alpha\beta\gamma I})^2
+e^{-1}{m\over48}{\epsilon}^{\alpha\beta\gamma\delta\epsilon\eta\xi}{\delta}^{IJ}S_{\alpha\beta\gamma
I}F_{\delta\epsilon\eta\xi J} 
\nonumber\\
&&  -{1\over2}P_{\alpha ij}P^{\alpha ij}
   +{{m^2}\over4}(T^2 -2T_{ij}T^{ij})
  -{{ie^{-1}\over{16\sqrt3}}}{\epsilon}^{\alpha\beta\gamma\delta\epsilon\eta\xi}
  {\epsilon}_{IKLMN}{\delta}^{IJ}S_{\alpha\beta\gamma
  J}F_{\delta\epsilon}^{KL}F_{\eta\xi}^{MN}
\nonumber\\
&&  +{e^{-1}\over{8m}}{\epsilon}^{\alpha\beta\gamma\delta\epsilon\eta\xi}Tr(
  B_{\alpha}F_{\beta\gamma}F_{\delta\epsilon}F_{\eta\xi} 
  -{4\over5}gB_{\alpha}B_{\beta}B_{\gamma}F_{\delta\epsilon}F_{\eta\xi}
  -{2\over5}gB_{\alpha}B_{\beta}F_{\gamma\delta}B_{\epsilon}F_{\eta\xi} 
\nonumber\\
&& {4\over5}{g^2}B_{\alpha}B_{\beta}B_{\gamma}B_{\delta}B_{\epsilon}F_{\eta\xi}
  -{8\over35}{g^3}B_{\alpha}B_{\beta}B_{\gamma}B_{\delta}B_{\epsilon}B_{\eta}
  B_{\xi}) 
\nonumber\\
&&  -{e^{-1}\over{16m}}{\epsilon}^{\alpha\beta\gamma\delta\epsilon\eta\xi}Tr(
  B_{\alpha}F_{\beta\gamma}-{2\over3}gB_{\alpha}B_{\beta}B_{\gamma})Tr(
  F_{\delta\epsilon}F_{\eta\xi}) 
\label{lagr7}
\eea

where the various field strengths are defined in terms of their
potentials as follows 
\bea
&  P_{\alpha ij}= {\Pi}^{-1 \; I}_{\; \; \; \; (i}({\delta}_{I}^{\; \; J}
  {\partial}_{\alpha}+2mB_{\alpha I}^{\; \; \; \; J}){\Pi}_{J}^{\; \; k}
  {\delta}_{j)k} 
\nonumber\\ 
&  T_{ij}={\Pi}^{-1 \; I}_{\; \; \; \; i}{\Pi}^{-1 \; J}_{\; \; \; \; j}
  {\delta}_{IJ} \; \; \; \; \; \; \; \; \; T=T_{ij}{\delta}^{ij} 
\nonumber\\
&  F_{\alpha\beta}^{\; \; \; \; IJ}={\delta}^{IK}
  F_{\alpha\beta K}^{\; \; \; \; \; \; J} 
\nonumber\\
&  F_{\alpha\beta I}^{\; \; \; \; \; \; J}=2({\partial}_{[\alpha}^{\  }
  B_{\beta ]I}^{\; \; \; \; \; J}+gB_{[\alpha I}^{\; \; \; \; \; K}
  B_{\beta ]K}^{\; \; \; \; \; J}) 
\nonumber\\
&  F_{\alpha\beta\gamma\delta I}=4({\partial}_{[\alpha}^{\ }
  S_{\beta\gamma\delta ]I}
  +gB_{[\alpha I}^{\; \; \; \; \; J}S_{\beta\gamma\delta ]J}).
\eea
The parameter $m$ is given in terms of the SO(5) coupling constant $g$
as $g=2m$. $I,J \dots = 1,{\dots},5$ are SO(5) vector indices and   
$i,j \dots = 1,{\dots},5$ are also vector indices but of a different
local composite SO(5)$_c$ group whose origin lies in the local Lorentz group in
eleven dimensions. The scalars ${\Pi}_{I}^{\; \; i}$ parametrise
the coset space SL(5,R)/SO(5)$_c$.
$B_{\alpha}^{\; \; IJ}= - B_{\alpha}^{\; \; JI}$
and we use the mostly plus metric convention.

To perform the reduction to D=6 the ansatz for the fields are:
\be
{\hat e}^{\; \hat a}_{\; \; \hat {\alpha}} = \pmatrix{
e^{\sigma\over2\sqrt{10}}
e^{\; \underline{\mu}}_{\; \; {\mu}} &0\cr
e^{-{2\sigma\over\sqrt{10}}} A_{\mu} & e^{-{2\sigma\over\sqrt{10}}}}
\label{5}
\ee 
where hats refer to D=7.
$\hat a =(\underline{\mu},\underline z)$ are local Lorenz indices and
$\hat {\alpha} =(\mu,z)$ are world indices.
\be 
d{\hat S_{7}}^{2}=e^{\sigma\over\sqrt{10}}dS_{6}^{2}+e^{-{4\sigma\over\sqrt{10}}} 
(dz+A)^{2}
\ee
where $A=A_{\mu}dx^{\mu}$ and $f=dA.$
\bea
&       {\hat {\Pi}_{I}^{\; \; i}}(x^{\mu},z)={\Pi}_{I}^{\; \;
       i}(x^{\mu}) 
\nonumber\\
&      {\hat B_{1 I}^{\; \; \; J}}=B_{1 I}^{\; \; \; J}+B_{0 I}^{\; \; \; J} 
      (dz+A) 
\nonumber\\
&      {\hat S_{3 I}^{\; \; \; J}}=S_{3 I}^{\; \; \; J}+S_{2 I}^{\; \; \; J} 
      (dz+A).
\eea
The resulting six dimensional lagrangian is:
\bea
  e^{-1}{\cal L}_6 &=& {R\over2}-{1\over8}e^{-{5\sigma\over\sqrt{10}}}
(f_{\mu\nu})^2
  -{1\over4}e^{-{\sigma\over\sqrt{10}}}({\Pi}_{I}^{\; \; i}{\Pi}_{J}^{\;
  \; j} F_{\mu\nu}^{IJ})^2
  -{{m^2}\over2}e^{-{2\sigma\over\sqrt{10}}}({\Pi}^{-1 \; I}_{\; \; \;
  \; i} S_{\mu\nu\rho I})^2 
\nonumber\\  
&&  -{3{m^2}\over2}e^{3\sigma\over\sqrt{10}}({\Pi}^{-1 \; I}_{\; \; \; \;
  i} S_{\mu\nu I})^2 
  -{1\over2}P_{\mu ij}P^{\mu
  ij}-{1\over2}e^{4\sigma\over\sqrt{10}}({\Pi}_{I}^{\; \;
  i}{\Pi}_{J}^{\; \; j} F_{\mu}^{IJ})^2
  -{1\over4}({\partial}_{\mu}{\sigma})^{2} 
\nonumber\\
&&    -{{m^2}\over2}e^{5\sigma\over\sqrt{10}}({\Pi}^{-1 \; \; I}_{\; \;
  \; \; (i} B_{0 I}^{\; \; \; J}{\Pi}_{J}^{\; \;
  k}{\delta}_{j)k})^2 
  +{{m^2}\over4}e^{\sigma\over\sqrt{10}}(T^2 -2T_{ij}T^{ij})
\nonumber\\
&&  +e^{-1}{m\over12}{\delta}^{IJ}{\epsilon}^{\mu\nu\rho\sigma\lambda\tau} \big [
  S_{\mu\nu\rho I}G_{\sigma\lambda\tau J}-gS_{\mu\nu\rho I}B_{0 J}^{\; \; \; K}
  S_{\sigma\lambda\tau K}
\nonumber\\
&&  +{3\over4}S_{\mu\nu I}G_{\rho\sigma\lambda\tau J}+{9\over2}S_{\mu\nu I} 
  S_{\rho\sigma J}{\partial}_{\lambda}A_{\tau} \big ]
  + {{e^{-1}}\over m}{\Omega(B)}
\nonumber\\
&&  -e^{-1}{{\sqrt{3}i}\over12}{\epsilon}_{JKLMN}{\delta}^{IJ}
  {\epsilon}^{\mu\nu\rho\sigma\lambda\tau} \big [ {3\over4}S_{\mu\nu I}
  F_{\rho\sigma}^{KL}F_{\lambda\tau}^{MN}+S_{\mu\nu\rho
  I}F_{\sigma\lambda}^{KL}F_{\tau}^{MN} \big ] 
\eea
where
\bea
&  P_{\mu ij}= {\Pi}^{-1 \; I}_{\; \; \; \; (i}({\delta}_{I}^{\; \; J}
  {\partial}_{\mu}+gB_{\mu I}^{\; \; \; \; J}){\Pi}_{J}^{\; \; k}
  {\delta}_{j)k}
\nonumber\\
&  T_{ij}={\Pi}^{-1 \; I}_{\; \; \; \; i}{\Pi}^{-1 \; J}_{\; \; \; \; j}
  {\delta}_{IJ} \; \; \; T=T_{ij}{\delta}^{ij} \; \; \; f_{2}=dA 
\nonumber\\
&   F_{2 I}^{\; \; \; J}= dB_{1 I}^{\; \; \; J}+B_{0 I}^{\; \; \;
   J}dA+gB_{1 I}^{\; \; \; K}B_{1 K}^{\; \; \; J} 
\nonumber\\
&   F_{1 I}^{\; \; \; J}= dB_{0 I}^{\; \; \; J}+g(B_{1 I}^{\; \;
   \; K}B_{0 K}^{\; \; \; J}-B_{0 I}^{\; \; \; K} B_{1 K}^{\; \; \; J})
\nonumber\\
&  G_{3 I}=dS_{2 I}+gB_{1 I}^{\; \; \; J}S_{2 J} \; \; \; \; 
  G_{4 I}=dS_{3 I}+gB_{1 I}^{\; \; \; J}S_{3 J} 
\eea
and the Chern-Simons term ${\Omega(B)}$ is given in appendix B.

After integrating by parts and dropping a total derivative we find 
\be
  {m\over16}{\epsilon}^{\mu\nu\rho\sigma\lambda\tau} 
  S_{\mu\nu}^{\; \; \; \; I}G_{\rho\sigma\lambda\tau I}
  ={m\over12}{\epsilon}^{\mu\nu\rho\sigma\lambda\tau} 
  S_{\mu\nu\rho}^{\; \; \; \; \; \; I}G_{\sigma\lambda\tau I} 
\ee
so the terms in the lagrangian involving $S_{2I}$ and $S_{3I}$ can be
compactly written as \cite{PKT}
\bea
 {\cal L}_{6}^{'} &=& {m^2}S_{\mu\nu I}Q^{\mu\nu\rho\sigma IJ}S_{\rho\sigma J}
  -{m^2}S_{\mu\nu\rho I}P^{\mu\nu\rho\sigma\lambda\tau
  IJ}S_{\sigma\lambda\tau J} 
\nonumber\\
&&  -{{\sqrt{3}i}\over12}{\epsilon}_{JKLMN}
  {\epsilon}^{\mu\nu\rho\sigma\lambda\tau}S_{\mu\nu }^{\; \; \; J}
  F_{\rho\sigma}^{KL}F_{\lambda\tau}^{MN} 
\nonumber\\
&&  +{m\over6}{\delta}^{IJ}{\epsilon}^{\mu\nu\rho\sigma\lambda\tau}
  S_{\mu\nu\rho I} \big (G_{\sigma\lambda\tau J}-{{\sqrt{3}i}\over2m} 
  {\epsilon}_{JKLMN}F_{\sigma\lambda}^{KL}F_{\tau}^{MN})
\eea
where the operators $P$ and $Q$ are defined by
\bea
  P^{\mu\nu\rho\sigma\lambda\tau IJ} &=& {e\over2}
  e^{-{2\sigma\over\sqrt{10}}}  
  {\Pi}^{-1 \; I}_{\; \; \; \; i} {\Pi}^{-1 \; iJ}g^{\mu\sigma}
  g^{\nu\lambda}g^{\rho\tau}+{1\over6}{\delta}^{IK}
  {\epsilon}^{\mu\nu\rho\sigma\lambda\tau}B_{0 K}^{\; \; \; J} 
\nonumber\\
  Q^{\mu\nu\rho\sigma IJ} &=& -{3e\over2}e^{3\sigma\over\sqrt{10}}  
  {\Pi}^{-1 \; I}_{\; \; \; \; i} {\Pi}^{-1 \; iJ}g^{\mu\rho}
  g^{\nu\sigma}+{3\over8m}{\delta}^{IJ}
  {\epsilon}^{\mu\nu\rho\sigma\lambda\tau}{\partial}_{\lambda}A_{\tau}.
\label{12}
\eea We have seen that the 3-index antisymmetric tensor potential in
D=7 splits into a 2-index and another 3-index tensor on reduction to
D=6. The D=7 odd dimensional self duality then allows the elimination
of the 3-index tensor in favour of the 2-index tensor. Variation of
${\cal L}_{6}^{'}$ w.r.t. $S_{3I}$ gives the equation \be
P^{\mu\nu\rho\sigma\lambda\tau IJ}S_{\sigma\lambda\tau J}=
{{1\over12m}}{\delta}^{IJ}{\epsilon}^{\mu\nu\rho\sigma\lambda\tau}
\big (G_{\sigma\lambda\tau J}-{{\sqrt{3}i}\over2m}
{\epsilon}_{JKLMN}F_{\sigma\lambda}^{KL}F_{\tau}^{MN}).  \ee Defining
the inverse of $P$ such that \be P^{\mu\nu\rho\alpha\beta\gamma
IJ}(P^{-1})_{\alpha\beta\gamma\sigma \lambda\tau JK}=
{\delta}_{\sigma\lambda\tau}^{\mu\nu\rho} {\delta}_{K}^{I} \ee and
using $P^{\mu\nu\rho\sigma\lambda\tau IJ}=
P^{\sigma\lambda\tau\mu\nu\rho JI}$ we have \be S_{\sigma\lambda\tau
P}={{1\over12m}}{\delta}^{IJ} {\epsilon}^{\alpha\beta\gamma\mu\nu\rho}
\big (G_{\mu\nu\rho I}-{{\sqrt{3}i}\over2m}
{\epsilon}_{IKLMN}F_{\mu\nu}^{KL}F_{\rho}^{MN})
P_{\alpha\beta\gamma\sigma\lambda\tau JP}^{-1}.  \ee Substitution back
in ${\cal L}_{6}^{'}$ finally yields the bosonic sector of the SO(5)
gauged D=6 model : \bea {\cal L}_6 &=&
{eR\over2}-{e\over8}e^{-{5\sigma\over\sqrt{10}}} (f_{\mu\nu})^2
-{e\over4}e^{-{\sigma\over\sqrt{10}}}({\Pi}_{I}^{\; \; i}{\Pi}_{J}^{\;
\; j} F_{\mu\nu}^{IJ})^2 +{m^2}S_{\mu\nu I}Q^{\mu\nu\rho\sigma
IJ}S_{\rho\sigma J} \nonumber\\ && +{1\over144} \big
[{\delta}^{I'J'}{\epsilon}^{\alpha\beta\gamma\delta
\epsilon\eta}({\tilde G}_{\delta\epsilon\eta I'}) \big ]
P_{\alpha\beta\gamma\mu\nu\rho J'I}^{-1} \big
[{\delta}^{IJ}{\epsilon}^{\mu\nu\rho\sigma\lambda\tau} ({\tilde
G}_{\sigma\lambda\tau J}) \big ] \nonumber\\ && -{e\over2}P_{\mu
ij}P^{\mu ij} -{e\over2}e^{4\sigma\over\sqrt{10}}({\Pi}_{I}^{\; \;
i}{\Pi}_{J}^{\; \; j} F_{\mu}^{IJ})^2
-{e\over4}({\partial}_{\mu}{\sigma})^{2}
-{e{m^2}\over2}e^{5\sigma\over\sqrt{10}}({\Pi}^{-1 \; \; I}_{\; \; \;
\; (i} B_{0 I}^{\; \; \; J}{\Pi}_{J}^{\; \; k}{\delta}_{j)k})^2
\nonumber\\ && +{e{m^2}\over4}e^{\sigma\over\sqrt{10}}(T^2
-2T_{ij}T^{ij}) -{\sqrt{3}i\over16}{\epsilon}_{JKLMN}
{\epsilon}^{\mu\nu\rho\sigma\lambda\tau}S_{\mu\nu }^{\; \; \; J}
F_{\rho\sigma}^{KL}F_{\lambda\tau}^{MN} + {1\over m}{\Omega(B)}
\label{16}
\eea where \bea & P_{\mu ij}= {\Pi}^{-1 \; I}_{\; \; \; \;
(i}({\delta}_{I}^{\; \; J} {\partial}_{\mu}+gB_{\mu I}^{\; \; \; \;
J}){\Pi}_{J}^{\; \; k} {\delta}_{j)k} \nonumber\\ & T_{ij}={\Pi}^{-1
\; I}_{\; \; \; \; i}{\Pi}^{-1 \; J}_{\; \; \; \; j} {\delta}_{IJ} \;
\; \; T=T_{ij}{\delta}^{ij} \; \; \; f_{2}=dA \nonumber\\ & F_{2
I}^{\; \; \; J}= dB_{1 I}^{\; \; \; J}+B_{0 I}^{\; \; \; J}dA+gB_{1
I}^{\; \; \; K}B_{1 K}^{\; \; \; J} \nonumber\\ & F_{1 I}^{\; \; \;
J}= dB_{0 I}^{\; \; \; J}+g(B_{1 I}^{\; \; \; K}B_{0 K}^{\; \; \;
J}-B_{0 I}^{\; \; \; K} B_{1 K}^{\; \; \; J}) \nonumber\\ & G_{3
I}=dS_{2 I}+gB_{1 I}^{\; \; \; J}S_{2 J} \nonumber\\ & {\tilde
G}_{\mu\nu\rho I} = G_{\mu\nu\rho I}-{{\sqrt{3}i}\over2m}
{\epsilon}_{IKLMN}F_{\mu\nu}^{KL}F_{\rho}^{MN} \eea and the operator
$Q$ is defined as in (\ref{12}). This model has a bosonic field
content of 1 graviton, (14+10+1) scalars, (10+1) vectors and 5 second
rank massive antisymmetric tensors. As all local supersymmetries of
the action remain unbroken after reduction on S$^1$, this SO(5) gauged
model is guaranteed a supersymmetric extension. In appendix A we
present the complete Lagrangian and supersymmetry transformation laws
of this D=6 SO(5) supergravity.\footnote{We note that the
transformation law of the auxiliary D=6 field S$_{3 I}$ becomes, upon
its elimination via the field equations, a consistency condition which
we expect to be satisfied upon use of the field equations and
supersymmetry transformation laws although we have not shown
this. Taking the $g=0$ limit and performing a dualisation, this
consistency condition becomes the supersymmetry transformation law of
the field $H_{\mu\nu\rho I}$.}

\section{The $g{\rightarrow}0$ limit}

In this section we show that the parameter $g$ can be taken to zero
smoothly. In order to show this it is necessary to rescale the various
fields and to regurgitate five vectors from the five 2-index
potentials $S_{2I}$. Each of these five second rank antisymmetric
tensor potentials is massive and so propagates 10 degrees of
freedom. On taking the $g{\rightarrow}0$ limit they become massless and so
propagate only 6 degrees of freedom. Hence the extra four degrees of
freedom from each $S_{2I}$ must form a massless vector. Therefore
$S_{\mu\nu I}$ is replaced everywhere by
\be
S_{\mu\nu I}-{1\over{m^2}}G_{\mu\nu I}
\ee
where $G_{\mu\nu I}=2{\partial}_{[\mu}S_{\nu ]I}$. This is
similar to the procedure used to formally recover the massless N=2a
D=10 supergravity from the massive model of Romans \cite{romans1}.

To obtain a lagrangian in a form from which the $g{\rightarrow}0$ limit can be
taken, the following field rescalings are necessary
\bea
&  B_{1 I}^{\; \; \; J}{\longrightarrow}g{B}_{1 I}^{\; \; \; J}
  {\qquad\qquad}{B}_{0 I}^{\; \; \; J}{\longrightarrow}g
  {B}_{0 I}^{\; \; \; J} 
\nonumber\\
& \qquad
  {\Pi}_{I}^{\; \; i}{\longrightarrow}{1\over\sqrt{g}}\;{\Pi}_{I}^{\; \; i}
  {\qquad\; \; \; \; \; }{\Pi}^{-1 \; I}_{\; \; \; \; i}
  {\longrightarrow}{\sqrt{g}}
  \;{\Pi}^{-1 \; I}_{\; \; \; \; i} 
\nonumber\\
&  S_{2 I}{\longrightarrow}{\sqrt{g}}\;S_{2 I}
  {\qquad\qquad}S_{1 I}{\longrightarrow}{\sqrt{g}}\;S_{1 I}.
\label{19}
\eea
These rescalings cause slight redefinitions in the field strengths. 
Notice that the operator $P$ scales with a factor of $g$. The
lagrangian now becomes

\bea
  {\cal L}_6 &=& {eR\over2}-{e\over8}e^{-{5\sigma\over\sqrt{10}}}
(f_{\mu\nu})^2
  -{e\over4}e^{-{\sigma\over\sqrt{10}}}({\Pi}_{I}^{\; \; i}{\Pi}_{J}^{\;
  \; j} F_{\mu\nu}^{IJ})^2 -6e{m^4}e^{3\sigma\over\sqrt{10}}S_{\mu\nu
  I} {\Pi}^{-1 \; I}_{\; \; \; \; i}{\Pi}^{-1 \; iJ}S_{\; \; \; \;
  J}^{\mu\nu}
\nonumber\\
&&  {3{m^2}\over4}{\delta}^{IJ}{\epsilon}^{\mu\nu\rho\sigma\lambda\tau}
  S_{\mu\nu I}S_{\rho\sigma J}{\partial}_{\lambda}A_{\tau}
  +12e{m^2}e^{3\sigma\over\sqrt{10}}S_{\mu\nu I}
  {\Pi}^{-1 \; I}_{\; \; \; \; i}{\Pi}^{-1 \; iJ}G_{\; \; \; \;
  J}^{\mu\nu} 
\nonumber\\
&&  -{3\over2}{\delta}^{IJ}{\epsilon}^{\mu\nu\rho\sigma\lambda\tau}
  S_{\mu\nu I}G_{\rho\sigma J}{\partial}_{\lambda}A_{\tau}
  -6ee^{3\sigma\over\sqrt{10}}G_{\mu\nu I}
  {\Pi}^{-1 \; I}_{\; \; \; \; i}{\Pi}^{-1 \; iJ}G_{\; \; \; \;
  J}^{\mu\nu}
\nonumber\\
&&  +{3\over{4{m^2}}}{\delta}^{IJ}{\epsilon}^{\mu\nu\rho\sigma\lambda\tau}
  G_{\mu\nu I}G_{\rho\sigma J}{\partial}_{\lambda}A_{\tau} 
\nonumber\\
&&  +{1\over144} \big [{\epsilon}^{\alpha\beta\gamma\delta
  \epsilon\eta}({\tilde G}_{\delta\epsilon\eta}^{\; \; \; \; \; J'}) \big ]
  P_{\alpha\beta\gamma\mu\nu\rho J'I}^{-1}
  \big [{\epsilon}^{\mu\nu\rho\sigma\lambda\tau}
  ({\tilde G}_{\sigma\lambda\tau}^{\; \; \; \; \; I}) \big ] 
\nonumber\\
&&  -{e\over2}P_{\mu ij}P^{\mu ij}
  -{e\over2}e^{4\sigma\over\sqrt{10}}({\Pi}_{I}^{\; \; i}{\Pi}_{J}^{\; \;
  j} F_{\mu}^{IJ})^2 
  -{e\over4}({\partial}_{\mu}{\sigma})^{2} 
\nonumber\\
&&    -{2e{m^4}}e^{5\sigma\over\sqrt{10}}({\Pi}^{-1 \; \; I}_{\; \; \; \;
  (i} B_{0 I}^{\; \; \; J}{\Pi}_{J}^{\; \; k}{\delta}_{j)k})^2 
  +{e{m^4}}e^{\sigma\over\sqrt{10}}(T^2 -2T_{ij}T^{ij})
\nonumber\\
&&  -{{\sqrt{6}i}\over4}{m^{5\over2}}{\epsilon}_{JKLMN}
  {\epsilon}^{\mu\nu\rho\sigma\lambda\tau}S_{\mu\nu}^{\; \; \; \; J}
  F_{\rho\sigma}^{KL}F_{\lambda\tau}^{MN} 
\nonumber\\
&&  +{{\sqrt{6}i}\over4}{m^{1\over2}}{\epsilon}_{JKLMN}
  {\epsilon}^{\mu\nu\rho\sigma\lambda\tau}G_{\mu\nu}^{\; \; \; \; J}
  F_{\rho\sigma}^{KL}F_{\lambda\tau}^{MN} 
  +{m^3}{\Omega (B)}
\eea
where
\bea
&  P_{\mu ij}= {\Pi}^{-1 \; I}_{\; \; \; \; (i}({\delta}_{I}^{\; \; J}
  {\partial}_{\mu}+{g^2}B_{\mu I}^{\; \; \; \; J}){\Pi}_{J}^{\; \; k}
  {\delta}_{j)k}
\nonumber\\
&  F_{2 I}^{\; \; \; J}= dB_{1 I}^{\; \; \; J}+B_{0 I}^{\; \; \;
   J}dA+{g^2}B_{1 I}^{\; \; \; K}B_{1 K}^{\; \; \; J} 
\nonumber\\
&   F_{1 I}^{\; \; \; J}= dB_{0 I}^{\; \; \; J}+{g^2}(B_{1 I}^{\; \;
   \; K}B_{0 K}^{\; \; \; J}-B_{0 I}^{\; \; \; K} B_{1 K}^{\; \; \;
  J}) 
\nonumber\\
&  G_{3 I}=dS_{2 I}+{g^2}B_{1 I}^{\; \; \; J}S_{2 J} \; \; \; \; \; 
  G_{2 I}=dS_{1 I} 
\nonumber\\
&  {\tilde G}_{\mu\nu\rho I} = G_{\mu\nu\rho I}-{2\sqrt{3m}i}
  {\epsilon}_{IKLMN}F_{\mu\nu}^{KL}F_{\rho}^{MN}
  -12B_{\mu I}^{\; \; \; J}{G}_{\nu\rho J}
\eea
\noindent
and ${\Omega(B)}$ is given in appendix B.

Noting that the term with an inverse power of $m$ is a total
derivative and so can be dropped, the $g{\rightarrow}0$ limit
can now be recovered
\bea
  {\cal L}_6 &=& {eR\over2}-{e\over8}e^{-{5\sigma\over\sqrt{10}}}
(f_{\mu\nu})^2
  -{e\over4}e^{-{\sigma\over\sqrt{10}}}({\Pi}_{I}^{\; \; i}{\Pi}_{J}^{\;
  \; j} F_{\mu\nu}^{IJ})^2 -6ee^{3\sigma\over\sqrt{10}}({\Pi}^{-1 \;
  I}_{\; \; \; \; i}G_{\mu\nu I})^{2} 
\nonumber\\
&&  +{1\over144} \big [{\epsilon}^{\alpha\beta\gamma\delta
  \epsilon\eta}({G}_{\delta\epsilon\eta}^{\; \; \; \; \; I}) \big ]
  P_{\alpha\beta\gamma\mu\nu\rho IJ}^{-1}
  \big [{\epsilon}^{\mu\nu\rho\sigma\lambda\tau}
  ({G}_{\sigma\lambda\tau}^{\; \; \; \; \; J}) \big ]
  -{e\over2}P_{\mu ij}P^{\mu ij} 
\nonumber\\
&&  -{e\over2}e^{4\sigma\over\sqrt{10}}({\Pi}_{I}^{\; \; i}{\Pi}_{J}^{\;
  \; j} F_{\mu}^{IJ})^2
  -{e\over4}({\partial}_{\mu}{\sigma})^{2}
  -{3\over2}{\delta}^{IJ}{\epsilon}^{\mu\nu\rho\sigma\lambda\tau}
  S_{\mu\nu I}G_{\rho\sigma J}{\partial}_{\lambda}A_{\tau} 
\eea
where 
\bea
&  f_{2}=dA \; \; \; \; \; \; \; \; G_{2I}=dS_{1I}
\nonumber\\
&  P_{\mu ij}= {\Pi}^{-1 \; I}_{\; \; \; \; (i}
  {\partial}_{\mu}^{\ }{\Pi}_{Ij)}^{\ }
\nonumber\\
&  F_{2 I}^{\; \; \; J}= dB_{1 I}^{\; \; \; J}+B_{0 I}^{\; \; \;
   J}dA \; \; \; \; \; \; \; F_{1 I}^{\; \; \; J}= dB_{0 I}^{\; \; \; J}
\nonumber\\
&  {G}_{\mu\nu\rho I} = 3{\partial}_{[ \mu}S_{\nu\rho ]I}
  -12B_{[ \mu I}^{\; \; \; \; J}{G}_{\nu\rho ]J} 
\eea
and 
\be
  P^{\mu\nu\rho\sigma\lambda\tau IJ}= {e\over2}
  e^{-{2\sigma\over\sqrt{10}}}  
  {\Pi}^{-1 \; I}_{\; \; \; \; i} {\Pi}^{-1 \; iJ}g^{\mu\sigma}
  g^{\nu\lambda}g^{\rho\tau}+{1\over6}{\delta}^{IK}
  {\epsilon}^{\mu\nu\rho\sigma\lambda\tau}B_{0 K}^{\; \; \; \; J}. 
\label{24}
\ee
\noindent
In order to make contact with the maximal D=6 supergravity of Tanii
\cite{Tanii} the two form potential $S_{2I}$ must be dualised to another two
form potential $C_{2I}$. The relevant sector of the Lagrangian is
\be
  {\cal L}= {1\over144} \big [{\epsilon}^{\alpha\beta\gamma\delta
  \epsilon\eta}({G}_{\delta\epsilon\eta}^{\; \; \; \; \; I}) \big ]
  P_{\alpha\beta\gamma\mu\nu\rho IJ}^{-1} \big
  [{\epsilon}^{\mu\nu\rho\sigma\lambda\tau}
  ({G}_{\sigma\lambda\tau}^{\; \; \; \; \; J}) \big ] 
  +{3\over2}{\epsilon}^{\mu\nu\rho\sigma\lambda\tau}
  {\partial}_{\mu}S_{\nu\rho I}G_{\sigma\lambda}{}^{I}A_{\tau}.
\ee
\noindent
To dualise $S_{2I}$ we replace $3{\partial}_{\mu}S_{\nu\rho I}$ by the
independent field $a_{\mu\nu\rho I}$ and add to the Lagrangian the term:
\be
    \Delta {\cal L}
={\kappa}{\epsilon}^{\mu\nu\rho\sigma\lambda\tau}a_{\mu\nu\rho I}H_{\sigma\lambda\tau}{}^{I},
\ee 
\noindent
where $\kappa$ is a constant and
$H_{\sigma\lambda\tau}{}^{I}=3{\partial}_{\sigma}C_{\lambda\tau}{}^{I}$.
Variation of ${\cal L}+{\Delta}{\cal L}$ w.r.t. $a_{\mu\nu\rho I}$ gives
\be
    {1\over72}P^{-1}_{\sigma\lambda\tau\alpha\beta\gamma
        IJ}[{\epsilon}^{\alpha\beta\gamma\delta\epsilon\eta}(a_{\delta\epsilon\eta}{}^{J}-12B_{\delta}{}^{JK}G_{\epsilon\eta K})]={\kappa}{H}^{\prime}_{\sigma\lambda\tau I},
\ee
\noindent
where ${H}^{\prime}_{\sigma\lambda\tau I}={H}_{\sigma\lambda\tau
I}+{1\over{2\kappa}}G_{\sigma\lambda I}A_{\tau}$. Substituting back
into ${\cal L}+{\Delta}{\cal L}$ one finds
\be
        {\cal L}+{\Delta}{\cal L}=-{{{\kappa}^{2}}\over4}(144){H}^{\prime}_{\mu\nu\rho
        I}P^{\mu\nu\rho\sigma\lambda\tau IJ}{H}^{\prime}_{\sigma\lambda\tau J}-12{\kappa}{\epsilon}^{\mu\nu\rho\sigma\lambda\tau}{H}^{\prime}_{\mu\nu\rho
        I}B_{\sigma}{}^{IJ}G_{\lambda\tau J}.
\ee
\noindent
Thus using the expression for $P^{\mu\nu\rho\sigma\lambda\tau IJ}$
(\ref{24}), ${\cal L}+{\Delta}{\cal L}$ becomes
\bea
{\cal L}+{\Delta}{\cal L} &=&
        -18{\kappa}^{2}ee^{-{2\sigma\over\sqrt{10}}}{({\Pi}^{-1 \;
  I}_{\; \; \; \; i}{H}^{\prime}_{\mu\nu\rho
  I})}^{2}-6{\kappa}^{2}{\epsilon}^{\mu\nu\rho\sigma\lambda\tau}B_{0}{}^{IJ}{H}^{\prime}_{\mu\nu\rho
  I}{H}^{\prime}_{\sigma\lambda\tau J}
\nonumber\\
&&  -12{\kappa}{\epsilon}^{\mu\nu\rho\sigma\lambda\tau}{H}^{\prime}_{\mu\nu\rho
  I}B_{\sigma}{}^{IJ}G_{\lambda\tau J}.
\eea
\noindent
Hence all mention of $P^{-1}$ has disappeared. Therefore finally the
bosonic sector of the ungauged D=6 model is (after a simple rescaling
of some fields, dropping primes on ${H}^{\prime}_{3}$ and choosing
$\kappa ={1\over3}$):
\bea
  e^{-1}{\cal L}_6 &=& R-{1\over4}e^{-{5\sigma\over\sqrt{10}}}
(f_{\mu\nu})^2
  -{1\over4}e^{-{\sigma\over\sqrt{10}}}({\Pi}_{I}^{\; \; i}{\Pi}_{J}^{\;
  \; j} F_{\mu\nu}^{IJ})^2 -{1\over4}e^{3\sigma\over\sqrt{10}}({\Pi}^{-1
  \; I}_{\; \; \; \; i}G_{\mu\nu I})^{2}
\nonumber\\
&&  -{1\over12}e^{-{2\sigma\over\sqrt{10}}}({\Pi}^{-1 \; I}_{\; \; \; \;
  i}H_{\mu\nu\rho I})^{2} -P_{\mu ij}P^{\mu
  ij}-{1\over2}e^{4\sigma\over\sqrt{10}}({\Pi}_{I}^{\; \;
  i}{\Pi}_{J}^{\; \; j} F_{\mu}^{IJ})^2
  -{1\over2}({\partial}_{\mu}{\sigma})^{2} 
\nonumber\\
&&  -{{e^{-1}}\over{36\sqrt2}}{\epsilon}^{\mu\nu\rho\sigma\lambda\tau}B_{0}{}^{IJ}{H}_{\mu\nu\rho
  I}{H}_{\sigma\lambda\tau
  J}-{e^{-1}\over{6\sqrt2}}{\epsilon}^{\mu\nu\rho\sigma\lambda\tau}{H}_{\mu\nu\rho
  I}B_{\sigma}{}^{IJ}G_{\lambda\tau J}
\label{30}
\eea
where 
\bea
&  f_{2}=dA \; \; \; \; \; \; \; \; G_{2I}=dS_{1I}
\nonumber\\
&  P_{\mu ij}= {\Pi}^{-1 \; I}_{\; \; \; \; (i}
  {\partial}_{\mu}^{\ }{\Pi}_{Ij)}^{\ }
\nonumber\\
&  F_{2 I}^{\; \; \; J}= dB_{1 I}^{\; \; \; J}+B_{0 I}^{\; \; \;
   J}dA \; \; \; \; \; \; \; F_{1 I}^{\; \; \; J}= dB_{0 I}^{\; \; \;
J} 
\nonumber\\
&  {H}_{\mu\nu\rho I} = 3({\partial}_{[ \mu}C_{\nu\rho ]I}
  +{1\over2}{G}_{[ \mu\nu I}{A}_{\rho ]}).
\eea
\noindent
The bosonic field content of this model is 1 graviton, (14+10+1)
scalars, (10+5+1) vectors and 5 second rank antisymmetric
tensors. This is precisely the bosonic field content of the maximally
extended supergravity in six dimensions \cite{Tanii}. Comparison of
the interactions is complicated by the manifestly SO(5,5) invariant
formulation of the ungauged model \cite{Tanii} but the interaction
terms in (\ref{30}) seem to have at least qualitatively the correct
structure. As all local supersymmetries of the action remain unbroken
after S$^1$ reduction, the Lagrangian (\ref{30}) must have a
supersymmetric extension. In appendix C we present the complete
Lagrangian and supersymmetry transformation laws of this D=6 maximal
supergravity.

As this section has been concerned with limits in which gauge
coupling constants in gauged supergravities are turned off, 
we take this opportunity to note that contrary
to the statement in \cite{romans2}, the limit in which the U(1)
coupling constant of D=5 N=4 SU(2)$\times$U(1) gauged supergravity is
turned off, can be recovered. This results in the D=5 N=4 SU(2) gauged
model obtained by the reduction on T$^2$ of D=7 N=2 SU(2) gauged
supergravity \cite{PMC,TvN}. In appendix D we verify this statement.

\section{Conclusion}

We have reduced to six dimensions the maximal D=7 SO(5) gauged
supergravity \cite{PPvN} giving details of the reduction of the
bosonic sector only. After eliminating some auxiliary fields one
obtains a new SO(5) gauged supergravity whose field content is 1 graviton, 5
massive 2-index tensor potentials, 11 vectors, 25 scalars, 4
gravitini and 20 spin 1/2 fermions. As all local supersymmetries of
the action remain unbroken after reduction on S$^1$ this reduced D=6
model must have a supersymmetric extension which we have presented in
an appendix. Since the D=7 SO(5) gauged supergravity can be obtained
from eleven dimensional supergravity by compactification on S$^4$,
this new D=6 SO(5) gauged supergravity is then the compactification on
S$^4{\times}$S$^1$ of D=11 supergravity. Performing this reduction
with the S$^1$ and S$^4$ factors in reverse order must lead to the
same D=6 SO(5) gauged model. Therefore the reduction of D=10 N=2a
supergravity on S$^4$ must also yield the D=6 SO(5) gauged model
presented in this paper, and it would be interesting to verify
this. The existence of an S$^4$ compactification of D=10 IIa
supergravity was also concluded in \cite{BST} through the discovery of
a dual frame in which the near horizon geometry of the D4-brane
supergravity solution is AdS$_6\times$S$^4$. Having obtained a new D=6
SO(5) gauged supergravity, we note that one may also perform further
S$^1$ reductions to yield new gauged maximal supergravities in D=5 and
D=4.

It is interesting that, unlike its seven dimensional progenitor, our
new D=6 SO(5) gauged supergravity has no `gauge discontinuity'. We
have shown that after regurgitating five vectors from the 2-index
tensor potentials, a procedure which is essentially the reverse of the
Higgs mechanism, one can take the $g{\rightarrow}0$ limit in the
bosonic sector. The bosonic field rescalings (\ref{19}) necessary to
show this are also sufficient to ensure the $g=0$ limit is obtainable
without problems in the fermionic sector and supersymmetry
transformation laws.

The resulting model has a field content of 1 graviton, 5 massless
2-index tensor potentials, 16 vectors, 25 scalars, 4 gravitini and 20
spin 1/2 fermions and is presumably equivalent to the maximally
extended supergravity of Tanii \cite{Tanii} in which all the internal
symmetries are manifest.

As shown in \cite{PPvNW} there exist non-compact maximal gauged supergravities
in D=7 with gauge groups SO(4,1) and SO(3,2). These models are simply
obtained from the SO(5) model by replacing all gauge fields by SO(p,q)
gauge fields and all ${\delta}^{IJ}$ contractions  with ${\eta}^{IJ}$
contractions where ${\eta}^{IJ}$=diag$(----+)$ or diag$(+++--)$ for
SO(4,1) or SO(3,2) respectively. Hence these models also have gauge
discontinuities and the results for the dimensionally reduced SO(5) 
gauged model are easily extended to these non-compactly gauged
models. The dimensionally reduced SO(4,1) model may also possess a
limit in D=6 where the gauging is only partially switched off and
SO(4,1) is contracted to ISO(4).

As mentioned in the introduction, the scalar potential of the D=7
SO(5) gauged supergravity has a supersymmetry preserving maximum which
is AdS$_7$ \cite{PPvNW}. It was pointed out in \cite{LPT} that Anti-De Sitter space
upon S$^1$ dimensional reduction gives a domain wall preserving 1/2
of the supersymmetry. Therefore one expects the D=6 SO(5) gauged model
(\ref{16}) to possess a 1/2 supersymmetric domain wall vacuum. We have
shown this to be the case by solving the D=6 Killing spinor
equations. Reduction
of the D=7 supersymmetry transformation law of the gravitino using the
ansatz (\ref{5}) with $A_{\mu}=0$ yields the D=6 Killing spinor equations:
\bea
  {\delta\chi} &=& {1\over2\sqrt{2}}{\tau}^{\mu}{\tau}^{\underline7}{\epsilon}
{\partial}_{\mu}{\sigma} 
-{5m\over4\sqrt{10}}{\tau}^{\underline7}{\epsilon}e^{{\sigma}\over{2\sqrt{10}}}
\\  
{\delta {\psi}_{\mu}} &=& {\partial}_{\mu}\epsilon +
  {1\over4}{\omega}_{\mu\underline{\nu}\underline{\rho}}
  {\tau}^{\underline{\nu}\underline{\rho}}\epsilon
  -{5\sqrt{2}m\over{16}}{\tau}_{\mu}{\epsilon}
  e^{{\sigma}\over{2\sqrt{10}}}  
  +{1\over{2\sqrt{10}}}{\partial}_{\mu}{\sigma}{\epsilon} 
\eea
where $\mu =0, \dots ,5$, underlined indices refer to `flat' space and
${\tau}^{\underline{7}}$ is the product of all of the D=6 Dirac matrices. 
Provided ${\epsilon}=H^{1\over16}(y){\epsilon}_{0}$ where 
${\tau}_{\underline{y}}{\epsilon}_{0}={\pm}{\epsilon}_{0}$, these
equations are satisfied in the domain wall background 
\bea
&  ds^{2}= H^{-{5\over12}}dx^{\alpha}dx^{\beta}{\eta_{\alpha\beta}}
  +H^{-{25\over12}}dy^{2} 
\nonumber\\
& e^{\sigma} = H^{\sqrt{10}\over12} 
\eea
where $H=c+3{\sqrt2}m|y| \; $ with c a positive constant and $ \;
{\alpha,\beta}= 0,{\dots},4$. Hence this domain wall preserves 1/2 of
the supersymmetry. See refs \cite{LPT,LPSS,stelle,CLPST} for more details of domain wall solutions in higher dimensional supergravities.

Further gauged maximal supergravities can also be found from
dimensional reduction of those gauged supergravities in D=7 and D=8.
The truncation of these models to non-maximal gauged supergravities is
also interesting. For example, the S$^1$ reduction and truncation of
D=8 SU(2) gauged supergravity \cite{SS1} may yield the D=7 SU(2) gauged
simple supergravity \cite{TvN}. More interestingly, the Scherk-Schwarz
reduction of the D=8 model may provide a way of obtaining the version
of D=7 SU(2) gauged supergravity which has a topological mass term. In
addition to reductions on tori, one may obtain new gauged maximal
supergravities from reductions on spheres of those models in D=7 and
D=8. For example, it has recently been proposed \cite{BPS} that the further
reduction of D=8 SU(2) gauged supergravity on S$^3$ would yield a D=5
SU(2)$\times$SU(2) maximal supergravity. That these reductions,
leading to new gauged supergravities, are possible, can be attributed
to the existence of supersymmetric intersecting brane solutions in
D=11 and D=10 having near horizon geometries of the form
AdS$_{k}\times$S$^{l}\times$S$^{m}\times\bE^{n}$.

In appendix D we have also shown that the limit of the bosonic sector
of D=5 N=4 SU(2)$\times$U(1) gauged supergravity, in which the U(1)
coupling constant is turned off, is the D=5 N=4 SU(2) gauged model obtained in \cite{PMC}.

\bigskip
{\it Acknowledgements} The author takes enormous pleasure in thanking
Paul Townsend for constant encouragement and criticisms.

\newpage
\section{Appendix A}

Here we present the complete Lagrangian and supersymmetry
transformation laws of D=6 SO(5) gauged supergravity. The field
content is 1 graviton $g_{\mu\nu}$, 5 2-index massive antisymmetric tensor potentials
$S_{\mu\nu I}$, (10+1) vectors ($B_{\mu I}^{\; \; \; \; J}, A_{\mu}$),
(14+10+1) scalars (${\Pi}_{I}^{\; \; j}, B_{0I}^{\; \; \; \; J},
{\sigma}$), 4 gravitini ${\psi}_{\mu}$ and (16+4) spin 1/2 fermions
(${\lambda}_{i}, {\chi}$). ${\psi}_{\mu}$, ${\lambda}_{i}$ and ${\chi}$
are all D=6 USp(4) symplectic Majorana spinors and the spinor
inversion formula reads
\bea
{\bar\lambda}{\tau}^{{\alpha}_{1}\dots{\alpha}_{n}}
{\gamma}^{{i}_{1}\dots{i}_{r}}{\chi} = (-1)^{n(n+1)\over2}(-1)^{r(r-1)\over2}
{\bar\chi}{\tau}^{{\alpha}_{1}\dots{\alpha}_{n}}
{\gamma}^{{i}_{1}\dots{i}_{r}}{\lambda}
\nonumber
\eea
where ${\gamma}^{i}$ are the D=5 Dirac matrices. The Lagrangian of D=6
SO(5) supergravity, neglecting quartic fermion terms, is:

\bea e^{-1}{\cal L}_6 &=&
  R-{1\over4}e^{-{5\sigma\over\sqrt{10}}}(f_{\mu\nu})^2
  -{1\over4}e^{-{\sigma\over\sqrt{10}}}({\Pi}_{I}^{\; \;
  i}{\Pi}_{J}^{\; \; j} F_{\mu\nu}^{IJ})^2 -P_{\mu ij}P^{\mu ij}
  -{1\over2}e^{4\sigma\over\sqrt{10}}({\Pi}_{I}^{\; \; i}{\Pi}_{J}^{\;
  \; j} F_{\mu}^{IJ})^2 \nonumber\\ &&
  -{1\over2}({\partial}_{\mu}{\sigma})^{2}
  -m^{2}e^{5\sigma\over\sqrt{10}}Q_{(ij)}Q^{(ij)}
  +{m^2}e^{\sigma\over\sqrt{10}}(T^2 -2T_{ij}T^{ij})
  -3{m^2}e^{3\sigma\over\sqrt{10}}({\Pi}^{-1 \; \; I}_{\; \; \; \;
  i}S_{\mu\nu I})^2 \nonumber\\ && +
  {{3{e^{-1}}m}\over{4\sqrt2}}{\epsilon}^{\mu\nu\rho\sigma\lambda\tau}S_{\mu\nu
  }^{\; \; \; I}S_{\rho\sigma I}{\partial}_{\lambda}A_{\tau}
  -{\sqrt{3}i{e^{-1}}\over{2\sqrt2}}{\epsilon}_{JKLMN}
  {\epsilon}^{\mu\nu\rho\sigma\lambda\tau}S_{\mu\nu }^{\; \; \; J}
  F_{\rho\sigma}^{KL}F_{\lambda\tau}^{MN} \nonumber\\ &&
  +{e^{-1}\over144} \big [{\epsilon}^{\alpha\beta\gamma\delta
  \epsilon\eta}{\tilde G}_{\delta\epsilon\eta}^{\; \; \; \; \; \; I}
  \big ] P_{\alpha\beta\gamma\mu\nu\rho IJ}^{-1} \big
  [{\epsilon}^{\mu\nu\rho\sigma\lambda\tau} {\tilde
  G}_{\sigma\lambda\tau }^{\; \; \; \; \; \; J} \big ] +
  {{e^{-1}}\over{2\sqrt{2}m}}{\Omega(B)} \nonumber\\ &&
  -{\bar\psi}_{\mu}{\tau}^{\mu\nu\rho}{\nabla}_{\nu}{\psi}_{\rho}
  -{\bar\chi}{\tau}^{\mu}{\nabla}_{\mu}{\chi}
  -{\bar\lambda}^{i}{\tau}^{\mu}{\nabla}_{\mu}{\lambda}_{i}
  +{m\over{2\sqrt2}}T{\bar\psi}_{\mu}{\tau}^{\mu\nu}{\psi}_{\nu}
  e^{\sigma\over2\sqrt{10}} \nonumber\\ &&
  -{m\over{2\sqrt{10}}}T{\bar\psi}_{\mu}{\tau}^{\mu}{\tau}^{\underline7}{\chi}
  e^{\sigma\over2\sqrt{10}}
  -{9m\over{20\sqrt2}}T{\bar\chi}{\chi}e^{\sigma\over2\sqrt{10}}
  -{m\over{2\sqrt2}}[8T^{ij}-T{\delta}^{ij}]{\bar\lambda}_{i}{\lambda}_{j}
  e^{\sigma\over2\sqrt{10}} \nonumber\\ &&
  +{\sqrt2}mT^{ij}{\bar\lambda}_{i}{\gamma}_{j}{\tau}^{\mu}{\psi}_{\mu}
  e^{\sigma\over2\sqrt{10}}
  -{2m\over\sqrt{10}}T^{ij}{\bar\lambda}_{i}{\gamma}_{j}{\tau}^{\underline7}{\chi}
  e^{\sigma\over2\sqrt{10}} +
  {\bar\psi}_{\mu}{\tau}^{\nu}{\tau}^{\mu}{\gamma}^{i}{\lambda}^{j}P_{\nu
  ij} \nonumber\\ && +
  2m{\bar\psi}_{\mu}{\tau}^{\underline7}{\tau}^{\mu}{\gamma}^{i}{\lambda}^{j}Q_{(ij)}e^{5\sigma\over2\sqrt{10}}
  +
  {2\sqrt{5}}m{\bar\chi}{\gamma}^{i}{\lambda}^{j}Q_{(ij)}e^{5\sigma\over2\sqrt{10}}
  +
  {1\over{4\sqrt{10}}}{\bar\psi}_{\mu}g^{\mu\nu}{\tau}^{\lambda}{\psi}_{\nu}{\partial}_{\lambda}\sigma
  \nonumber\\ &&
  -{1\over{\sqrt{2}}}{\bar\psi}_{\mu}{\tau}^{\nu}{\tau}^{\mu}{\tau}^{\underline7}
  {\chi}{\partial}_{\nu}\sigma
  -{1\over{8}}{\bar\psi}_{\mu}[2{\tau}^{\mu\nu\rho}{\tau}^{\lambda}+
  {\tau}^{\mu\rho}{\tau}^{\nu\lambda}]{\tau}^{\underline7}{\psi}_{\rho}f_{\nu\lambda}e^{-{5\sigma\over2\sqrt{10}}}
  \nonumber\\ &&
  -{{\sqrt5}\over8}{\bar\psi}_{\mu}{\tau}^{\nu\lambda}{\tau}^{\mu}{\chi}f_{\nu\lambda}e^{-{5\sigma\over2\sqrt{10}}}
  -{3\over{16}}{\bar\chi}{\tau}^{\underline7}{\tau}^{\nu\lambda}{\chi}f_{\nu\lambda}e^{-{5\sigma\over2\sqrt{10}}}
  -{1\over{8}}{\bar\lambda}^{i}{\tau}^{\underline7}{\tau}^{\nu\lambda}{\lambda}_{i}
  f_{\nu\lambda}e^{-{5\sigma\over2\sqrt{10}}} \nonumber\\ &&
  +{1\over{8\sqrt2}}{\bar\psi}_{\mu}[{\tau}^{\mu\nu\rho\sigma}-2{g}^{\mu\nu}
  {g}^{\rho\sigma}]{\gamma}_{ij}{\psi}_{\sigma}{\Pi}_{I}^{\; \;
  i}{\Pi}_{J}^{\; \; j}F_{\nu\rho}^{IJ}e^{-{\sigma\over2\sqrt{10}}}
  \nonumber\\ &&
  +{1\over{8\sqrt{10}}}{\bar\psi}_{\mu}[2{g}^{\mu\nu}{\tau}^{\rho}+{\tau}^{\mu\nu\rho}]{\tau}^{\underline7}{\gamma}_{ij}{\chi}{\Pi}_{I}^{\;
  \; i}{\Pi}_{J}^{\; \; j}F_{\nu\rho}^{IJ}e^{-{\sigma\over2\sqrt{10}}}
  \nonumber\\ &&
  -{11\over{80\sqrt{2}}}{\bar\chi}{\tau}^{\nu\rho}{\gamma}_{ij}{\chi}{\Pi}_{I}^{\;
  \; i}{\Pi}_{J}^{\; \; j}F_{\nu\rho}^{IJ}e^{-{\sigma\over2\sqrt{10}}}
  +{1\over{2\sqrt{2}}}{\bar\psi}_{\mu}{\tau}^{\nu\rho}{\tau}^{\mu}{\gamma}_{i}{\lambda}_{j}{\Pi}_{I}^{\;
  \; i}{\Pi}_{J}^{\; \; j}F_{\nu\rho}^{IJ}
  e^{-{\sigma\over2\sqrt{10}}} \nonumber\\ &&
  -{1\over{2\sqrt{10}}}{\bar\chi}{\tau}^{\nu\rho}{\tau}^{\underline7}{\gamma}_{i}{\lambda}_{j}{\Pi}_{I}^{\;
  \; i}{\Pi}_{J}^{\; \; j}F_{\nu\rho}^{IJ}
  e^{-{\sigma\over2\sqrt{10}}}
  +{1\over{16\sqrt{2}}}{\bar\lambda}_{i}{\gamma}^{j}{\gamma}_{kl}{\gamma}^{i}{\tau}^{\nu\rho}{\lambda}_{j}{\Pi}_{I}^{\;
  \; k}{\Pi}_{J}^{\; \; l}F_{\nu\rho}^{IJ}
  e^{-{\sigma\over2\sqrt{10}}} \nonumber\\
  &&-{1\over{4\sqrt{2}}}{\bar\psi}_{\mu}{\tau}^{\mu\nu\rho}{\tau}^{\underline7}{\gamma}_{ij}{\psi}_{\rho}{\Pi}_{I}^{\;
  \; i}{\Pi}_{J}^{\; \; j}F_{\nu}^{IJ} e^{2\sigma\over\sqrt{10}}
  -{1\over{\sqrt{10}}}{\bar\psi}_{\mu}{\tau}^{\mu}{\tau}^{\nu}{\gamma}_{ij}{\chi}{\Pi}_{I}^{\;
  \; i}{\Pi}_{J}^{\; \; j}F_{\nu}^{IJ}e^{2\sigma\over\sqrt{10}}
  \nonumber\\ &&
  +{7\over{20\sqrt{2}}}{\bar\chi}{\tau}^{\underline7}{\tau}^{\nu}{\gamma}_{ij}{\chi}{\Pi}_{I}^{\;
  \; i}{\Pi}_{J}^{\; \; j}F_{\nu}^{IJ}e^{2\sigma\over\sqrt{10}}
  -{1\over{\sqrt{2}}}{\bar\psi}_{\mu}{\tau}^{\nu}{\tau}^{\mu}{\tau}^{\underline7}{\gamma}_{i}{\lambda}_{j}{\Pi}_{I}^{\;
  \; i}{\Pi}_{J}^{\; \; j}F_{\nu}^{IJ} e^{2\sigma\over\sqrt{10}}
  \nonumber\\ &&
  +{4\over{\sqrt{10}}}{\bar\chi}{\tau}^{\nu}{\gamma}_{i}{\lambda}_{j}{\Pi}_{I}^{\;
  \; i}{\Pi}_{J}^{\; \; j}F_{\nu}^{IJ}e^{2\sigma\over\sqrt{10}}
  +{1\over{8\sqrt{2}}}{\bar\lambda}_{i}{\gamma}^{j}{\gamma}_{kl}{\gamma}^{i}{\tau}^{\nu}{\tau}^{\underline7}{\lambda}_{j}{\Pi}_{I}^{\;
  \; k}{\Pi}_{J}^{\; \; l}F_{\nu}^{IJ}e^{2\sigma\over\sqrt{10}}
  \nonumber\\ &&
  -{im\sqrt{3}\over{4}}{\bar\psi}_{\mu}[{\tau}^{\mu\nu\rho\lambda}+
  {g}^{\mu\nu}{g}^{\rho\lambda}]{\tau}^{\underline7}{\gamma}^{i}
  {\psi}_{\lambda}{\Pi}^{-1 \; \; I}_{\; \; \; \; i}S_{\nu\rho I}
  e^{3\sigma\over2\sqrt{10}} \nonumber\\ &&
  -{3im\sqrt{3}\over{4\sqrt{5}}}{\bar\psi}_{\mu}[{\tau}^{\mu\nu\rho}-2
  {g}^{\mu\nu}{\tau}^{\rho}]{\gamma}^{i}{\chi}{\Pi}^{-1 \; \; I}_{\;
  \; \; \; i}S_{\nu\rho I}e^{3\sigma\over2\sqrt{10}} \nonumber\\ &&
  -{i\sqrt{3}\over{40}}{\bar\chi}{\tau}^{\underline7}{\tau}^{\nu\rho}
  {\gamma}^{i}{\chi}{\Pi}^{-1 \; \; I}_{\; \; \; \; i}S_{\nu\rho I}
  e^{3\sigma\over2\sqrt{10}}
  -{im\sqrt{3}\over{2}}{\bar\psi}_{\mu}[{\tau}^{\mu\nu\rho}-2{g}^{\mu\nu}
  {\tau}^{\rho}]{\tau}^{\underline7}{\lambda}^{i}{\Pi}^{-1 \; \; I}_
  {\; \; \; \; i}S_{\nu\rho I}e^{3\sigma\over2\sqrt{10}} \nonumber\\
  &&
  +{9im\over{2\sqrt{15}}}{\bar\chi}{\tau}^{\nu\rho}{\lambda}^{i}{\Pi}^{-1
  \; \; I}_{\; \; \; \; i}S_{\nu\rho I}e^{3\sigma\over2\sqrt{10}}
  -{im\sqrt{3}\over{4}}{\bar\lambda}^{j}{\tau}^{\nu\rho}{\tau}^{\underline7}
  {\gamma}^{i}{\lambda}_{j}{\Pi}^{-1 \; \; I}_{\; \; \; \; i}
  S_{\nu\rho I}e^{3\sigma\over2\sqrt{10}} \nonumber\\ &&
  +{m\over2}{\bar\psi}_{\mu}{\tau}^{\mu\nu}{\tau}^{\underline7}Q_{[ij]}
  {\gamma}^{ij}{\psi}_{\nu}e^{5\sigma\over2\sqrt{10}}
  +{m\sqrt{5}\over2}{\bar\psi}_{\mu}{\tau}^{\mu}Q_{[ij]}
  {\gamma}^{ij}{\chi}e^{5\sigma\over2\sqrt{10}}
  -{3m\over4}{\bar\chi}{\tau}^{\underline7}Q_{[ij]}
  {\gamma}^{ij}{\chi}e^{5\sigma\over2\sqrt{10}} \nonumber\\ &&
  -{m\over2}{\bar\lambda}^{k}{\tau}^{\underline7}Q_{[ij]}
  {\gamma}^{ij}{\lambda}_{k}e^{5\sigma\over2\sqrt{10}}
  -2m{\bar\lambda}^{i}{\tau}^{\underline7}Q_{[ij]}{\lambda}^{j}
  e^{5\sigma\over2\sqrt{10}} \nonumber\\ && +{i\over{24\sqrt{6}}}\bigg
  \{
  {1\over2}{\bar\psi}_{\sigma}[{\tau}^{\sigma\mu\nu\rho\lambda}+6{g}^{\sigma\mu}
  {\tau}^{\nu}{g}^{\rho\lambda}]{\gamma}^{i}{\psi}_{\lambda}
  +{1\over{\sqrt{5}}}{\bar\psi}_{\sigma}[{\tau}^{\sigma\mu\nu\rho}-3{g}^{\sigma\mu}
  {\tau}^{\nu\rho}]{\tau}^{\underline7}{\gamma}^{i}{\chi} \nonumber\\
  &&+{3\over{10}}{\bar\chi}{\tau}^{\mu\nu\rho}{\gamma}^{i}{\chi}
  -{\bar\psi}_{\sigma}[{\tau}^{\sigma\mu\nu\rho}-3{g}^{\sigma\mu}
  {\tau}^{\nu\rho}]{\lambda}^{i}
  -{2\over\sqrt{5}}{\bar\chi}{\tau}^{\underline7}{\tau}^{\mu\nu\rho}{\lambda}^{i}
  \nonumber\\
  &&-{1\over2}{\bar\lambda}^{j}{\tau}^{\mu\nu\rho}{\gamma}^{i}{\lambda}_{j}
  \bigg \} {\Pi}^{-1 \; \; I}_{\; \; \; \; i}
  P_{\mu\nu\rho\delta\epsilon\eta IJ}^{-1} \big
  [{\epsilon}^{\delta\epsilon\eta\alpha\beta\gamma} {\tilde
  G}_{\alpha\beta\gamma}^{\; \; \; \; \; \;J} \big ]
  e^{-{\sigma\over\sqrt{10}}}.  \eea 

where 
\bea 
& T_{ij}={\Pi}^{-1 \;
  I}_{\; \; \; \; i}{\Pi}^{-1 \; J}_{\; \; \; \; j} {\delta}_{IJ} \;
  \; \; T=T_{ij}{\delta}^{ij} \; \; \; f_{2}=dA 
\nonumber\\ 
& F_{2
  I}^{\; \; \; J}= dB_{1 I}^{\; \; \; J}+B_{0 I}^{\; \; \; J}dA+gB_{1
  I}^{\; \; \; K}B_{1 K}^{\; \; \; J} 
\nonumber\\ 
& F_{1 I}^{\; \; \;
  J}= dB_{0 I}^{\; \; \; J}+g(B_{1 I}^{\; \; \; K}B_{0 K}^{\; \; \;
  J}-B_{0 I}^{\; \; \; K} B_{1 K}^{\; \; \; J}) 
\nonumber\\ 
& G_{3
  I}=dS_{2 I}+gB_{1 I}^{\; \; \; J}S_{2 J} 
\nonumber\\ 
& {\tilde
  G}_{\mu\nu\rho I} = G_{\mu\nu\rho I}-{{\sqrt{3}i}\over2g}
  {\epsilon}_{IKLMN}F_{\mu\nu}^{KL}F_{\rho}^{MN} 
\nonumber\\ 
& P_{\mu
  ij}= {\Pi}^{-1 \; I}_{\; \; \; \; (i}({\delta}_{I}^{\; \; J}
  {\partial}_{\mu}+gB_{\mu I}^{\; \; \; \; J}){\Pi}_{J}^{\; \; k}
  {\delta}_{j)k} 
\nonumber\\ 
& Q_{\mu ij}= {\Pi}^{-1 \; I}_{\; \; \;
  \; [i}({\delta}_{I}^{\; \; J} {\partial}_{\mu}+gB_{\mu I}^{\; \; \;
  \; J}){\Pi}_{J}^{\; \; k} {\delta}_{j]k} 
\nonumber\\ 
& Q_{ij}=
  {\Pi}^{-1 \; I}_{\; \; \; \; i}B_{0I}^{\; \; \; \; J} {\Pi}_{J}^{\;
  \; k}{\delta}_{jk} 
\nonumber\\ 
& P^{\mu\nu\rho\sigma\lambda\tau IJ}
  = {e\over2} e^{-{2\sigma\over\sqrt{10}}} {\Pi}^{-1 \; I}_{\; \; \;
  \; i} {\Pi}^{-1 \; iJ}g^{\mu\sigma}
  g^{\nu\lambda}g^{\rho\tau}+{1\over6\sqrt{2}}{\delta}^{IK}
  {\epsilon}^{\mu\nu\rho\sigma\lambda\tau}B_{0 K}^{\; \; \; J}, 
\eea
  
and 

\bea {\nabla}_{\mu}{\xi}_{\nu i} = {\partial}_{\mu}{\xi}_{\nu
  i}+{1\over4} {\omega}_{\mu\underline{\nu}\underline{\rho}}
  {\tau}^{\underline{\nu}\underline{\rho}} {\xi}_{\nu
  i}+{\Gamma}_{\mu\nu}^{\rho}{\xi}_{\rho i}+ {1\over4}{Q}_{\mu
  jk}{\gamma}^{jk}{\xi}_{\nu i}+ {Q}_{\mu i}^{\; \; \; j}{\xi}_{\nu
  j}, 
\eea 
for a general vector spinor, ${\xi}_{\nu i}$, of both
  SO(5,1) and SO(5)$_{c}$.  ${\psi}_{\mu}$ and ${\chi}$ are
  SO(5)$_{c}$ spinors and ${\lambda}_{i}$ is an SO(5)$_{c}$ vector
  spinor.  The supersymmetry transformations of the fields neglecting
  terms cubic in fermion fields are:

\bea
 {\delta}{\psi}_{\mu} &=& {\nabla}_{\mu}{\epsilon} - {\sqrt{2}m\over16}T
{\tau}_{\mu}{\epsilon}e^{\sigma\over{2\sqrt{10}}}
+ {m\over8}{\tau}_{\mu}{\tau}^{\underline7}Q_{[ij]}{\gamma}^{ij}{\epsilon}
e^{5\sigma\over{2\sqrt{10}}}
-{1\over32}[{\tau}_{\mu}^{\; \; \nu\rho}-6{\delta}_{\mu}^{\nu}{\tau}^{\rho}]
{\tau}^{\underline7}{\epsilon}f_{\nu\rho}e^{-{5\sigma\over{2\sqrt{10}}}}
\nonumber\\
&& -{1\over{32\sqrt{2}}}[{\tau}_{\mu}^{\; \; \nu\rho}-6{\delta}_{\mu}^{\nu}{\tau}^{\rho}]{\gamma}_{ij}{\epsilon}{\Pi}_{I}^{\;
  \; i}{\Pi}_{J}^{\; \; j}F_{\nu\rho}^{IJ}
e^{-{\sigma\over{2\sqrt{10}}}}
+ {1\over{4\sqrt{2}}}{\tau}^{\underline7}{\gamma}_{ij}{\epsilon}{\Pi}_{I}^{\;
  \; i}{\Pi}_{J}^{\; \; j}F_{\mu}^{IJ}
e^{2\sigma\over{\sqrt{10}}}
\nonumber\\
&& + {1\over{2\sqrt{10}}}{\epsilon}{\partial}_{\mu}{\sigma}
+{im\sqrt3\over{16}}[{\tau}_{\mu}^{\; \; \nu\rho}-6{\delta}_{\mu}^{\nu}
{\tau}^{\rho}]{\tau}^{\underline7}{\gamma}^{i}{\epsilon}{\Pi}^{-1 \;
  \; I}_{\; \; \; \; i}S_{\nu\rho I}e^{3\sigma\over2\sqrt{10}}
\nonumber\\
&&+{i\over{96\sqrt{6}}}[{\tau}_{\mu}^{\; \; \nu\rho\sigma}-3{\delta}_{\mu}^{\nu}
{\tau}^{\rho\sigma}]{\gamma}^{i}{\epsilon}{\Pi}^{-1 \; \; I}_{\;
  \; \; \; i}P_{\nu\rho\sigma\alpha\beta\gamma IJ}^{-1}
\big [{\epsilon}^{\alpha\beta\gamma\delta\epsilon\eta}
{\tilde G}_{\delta\epsilon\eta}^{\; \; \; \; \; \;J} \big ]
e^{-{\sigma\over\sqrt{10}}},
\eea
\bea
{\delta}{\chi}  &=& -{\sqrt{5}\over{16}}{\tau}^{\mu\nu}{\epsilon}f_{\mu\nu}
e^{-{5\sigma\over{2\sqrt{10}}}}
+{1\over{2\sqrt{2}}}{\tau}^{\mu}{\tau}^{\underline7}{\epsilon}
{\partial}_{\mu}{\sigma}
-{m\over{4\sqrt{10}}}T{\tau}^{\underline7}{\epsilon}e^{\sigma\over{2\sqrt{10}}}
\nonumber\\
&&-{1\over{16\sqrt{10}}}{\tau}^{\mu\nu}{\tau}^{\underline7}{\gamma}_{ij}
{\epsilon}{\Pi}_{I}^{\; \; i}{\Pi}_{J}^{\; \; j}F_{\mu\nu}^{IJ}
e^{-{\sigma\over{2\sqrt{10}}}}
-{1\over{2\sqrt{10}}}{\tau}^{\mu}{\gamma}_{ij}{\epsilon}
{\Pi}_{I}^{\; \; i}{\Pi}_{J}^{\; \; j}F_{\mu}^{IJ}
e^{2\sigma\over{\sqrt{10}}}
\nonumber\\
&&-{9im\over{8\sqrt{15}}}{\tau}^{\mu\nu}{\gamma}^{i}{\epsilon}
{\Pi}^{-1 \; \; I}_{\; \; \; \; i}S_{\mu\nu I}e^{3\sigma\over2\sqrt{10}}
+{m\sqrt{5}\over{4}}Q_{[ij]}{\gamma}^{ij}{\epsilon}
e^{5\sigma\over{2\sqrt{10}}}
\nonumber\\
&&-{i\over{48\sqrt{30}}}{\tau}^{\mu\nu\rho}{\tau}^{\underline7}{\gamma}^{i}
{\epsilon}{\Pi}^{-1 \; \; I}_{\; \; \; \; i}
P_{\mu\nu\rho\alpha\beta\gamma IJ}^{-1}
\big [{\epsilon}^{\alpha\beta\gamma\delta\epsilon\eta}
{\tilde G}_{\delta\epsilon\eta}^{\; \; \; \; \; \;J} \big ]
e^{-{\sigma\over\sqrt{10}}},
\eea
\bea
{\delta}{\lambda}_{i} &=&
{1\over{16\sqrt{2}}}{\tau}^{\mu\nu}[{\gamma}_{kl}{\gamma}_{i}-{1\over5}
{\gamma}_{i}{\gamma}_{kl}]{\epsilon}{\Pi}_{I}^{\; \; k}{\Pi}_{J}^{\;
\; l}F_{\mu\nu}^{IJ}e^{-{\sigma\over{2\sqrt{10}}}}
+{1\over2}{\tau}^{\mu}{\gamma}^{j}{\epsilon}P_{\mu ij}
\nonumber\\
&&+{1\over{8\sqrt{2}}}{\tau}^{\mu}{\tau}^{\underline7}
[{\gamma}_{kl}{\gamma}_{i}-{1\over5}{\gamma}_{i}{\gamma}_{kl}]{\epsilon}
{\Pi}_{I}^{\; \; k}{\Pi}_{J}^{\; \; l}F_{\mu}^{IJ}e^{2\sigma\over{\sqrt{10}}}
+m{\tau}^{\underline7}{\gamma}^{j}{\epsilon}Q_{(ij)}
e^{5\sigma\over{2\sqrt{10}}}
\nonumber\\
&&+{3im\over{20\sqrt{3}}}{\tau}^{\mu\nu}{\tau}^{\underline7}[{\gamma}_{i}^{\;
j}-4{\delta}_{i}^{\; j}]{\epsilon}{\Pi}^{-1 \; \; I}_{\; \; \; \;
j}S_{\mu\nu I}e^{3\sigma\over2\sqrt{10}}+
{m\over\sqrt{2}}[T_{ij}-{1\over5}{\delta}_{ij}T]
{\gamma}^{j}{\epsilon}e^{\sigma\over{2\sqrt{10}}}
\nonumber\\
&&+{i\over{240\sqrt{6}}}{\tau}^{\mu\nu\rho}[{\gamma}_{i}^{\;
j}-4{\delta}_{i}^{\; j}]{\epsilon}{\Pi}^{-1 \; \; I}_{\; \; \; \; j}
P_{\mu\nu\rho\alpha\beta\gamma IJ}^{-1}
\big [{\epsilon}^{\alpha\beta\gamma\delta\epsilon\eta}
{\tilde G}_{\delta\epsilon\eta}^{\; \; \; \; \; \;J} \big ]
e^{-{\sigma\over\sqrt{10}}},
\eea
\bea
{\delta\sigma} = -{1\over\sqrt2}{\bar\epsilon}{\tau}^{\underline7}\chi
e^{\sigma\over\sqrt{10}}, \; \; \; \; \; \; \; \; \; \; \; \; \; 
{\Pi}^{-1 \; \; I}_{\; \; \; \; i}{\delta}
{\Pi}_{I}^{\; \; j}= {1\over4}[{\bar\epsilon}{\gamma}_{i}{\lambda}^{j}+
{\bar\epsilon}{\gamma}^{j}{\lambda}_{i}]e^{\sigma\over\sqrt{10}},
\eea
\bea
{\delta}A_{\mu} = {1\over2}{\bar\epsilon}{\tau}^{\underline7}{\psi}_{\mu}
e^{7\sigma\over2\sqrt{10}}+{1\over4\sqrt5}{\bar\epsilon}{\tau}_{\mu}\chi
e^{7\sigma\over2\sqrt{10}},
\eea
\bea
{\delta}{e^{\underline{\mu}}}^{\ }_{\mu}={1\over2}{\bar\epsilon}
{\tau}^{\underline{\mu}}{\psi}_{\mu}e^{\sigma\over\sqrt{10}}
+ {1\over\sqrt{5}}{\bar\epsilon}{\tau}^{\underline{\mu}}{\chi}A_{\mu}
e^{-{3\sigma\over2\sqrt{10}}}
+{1\over4\sqrt{5}}{\bar\epsilon}{\tau}_{\mu}^{\; \; \nu}
{e^{\underline{\mu}}}^{\ }_{\nu}{\tau}^{\underline7}{\chi}
e^{\sigma\over\sqrt{10}},
\eea
\bea
{\Pi}_{I}^{\; \; i}{\Pi}_{J}^{\; \; j}{\delta}B_{0}^{\; \; IJ} &=&
{1\over\sqrt{10}}{\bar\epsilon}{\gamma}^{ij}{\chi}e^{-{\sigma\over\sqrt{10}}}
-{1\over\sqrt{5}}{\bar\epsilon}{\tau}^{\mu}\chi{\Pi}_{I}^{\; \;
i}{\Pi}_{J}^{\; \; j}B_{\mu}^{\; \; IJ}e^{-{3\sigma\over2\sqrt{10}}} 
\nonumber\\
&&+{1\over4\sqrt{2}}{\bar\epsilon}{\tau}^{\underline7}{\gamma}^{k}{\gamma}^{ij}
{\lambda}_{k}e^{-{\sigma\over\sqrt{10}}},
\eea
\bea
{\Pi}_{I}^{\; \; i}{\Pi}_{J}^{\; \; j}{\delta}B_{\mu}^{\; \; IJ} &=&
{1\over2\sqrt{2}}{\bar\epsilon}{\gamma}^{ij}{\psi}_{\mu}
e^{3\sigma\over2\sqrt{10}}
+{1\over4\sqrt{10}}{\bar\epsilon}{\gamma}^{ij}{\tau}^{\underline7}{\tau}_{\mu}
{\chi}e^{3\sigma\over2\sqrt{10}}
+{1\over4\sqrt{2}}{\bar\epsilon}{\tau}_{\mu}{\gamma}^{k}{\gamma}^{ij}
{\lambda}_{k}e^{3\sigma\over2\sqrt{10}}
\nonumber\\
&&-{1\over2}{\bar\epsilon}{\tau}^{\underline7}{\psi}_{\mu}{\Pi}_{I}^{\; \; i}{\Pi}_{J}^{\; \; j}B_{0}^{\; \; IJ}e^{7\sigma\over{2\sqrt{10}}}
-{1\over4\sqrt{5}}{\bar\epsilon}{\tau}_{\mu}\chi{\Pi}_{I}^{\; \; i}{\Pi}_{J}^{\; \; j}B_{0}^{\; \; IJ}e^{7\sigma\over{2\sqrt{10}}}
\nonumber\\
&&+{1\over\sqrt{5}}{\bar\epsilon}{\tau}^{\nu}A_{\mu}\chi{\Pi}_{I}^{\; \; i}{\Pi}_{J}^{\; \; j}B_{\nu}^{\; \; IJ}e^{-{3\sigma\over{2\sqrt{10}}}},
\eea
\bea
{\delta}S_{\mu\nu I} &=& -{2\over\sqrt{5}}{\bar\epsilon}{\tau}^{\rho}{\chi}
A_{\mu}S_{\nu\rho I}e^{-{3\sigma\over{2\sqrt{10}}}}
+ {i\over2\sqrt{6}}{\delta}_{IJ}{\Pi}^{-1 \; \; J}_{\; \; \; \; i}
[{1\over\sqrt{5}}{\bar\epsilon}{\tau}_{\mu\nu}{\gamma}^{i}\chi 
-2{\bar\epsilon}{\tau}_{\mu}{\tau}^{\underline7}{\gamma}^{i}{\psi}_{\nu}
-{\bar\epsilon}{\tau}_{\mu\nu}{\tau}^{\underline7}{\lambda}^{i}]
\nonumber\\
&& -{i\over8\sqrt{6}m}{\Pi}_{I}^{\; \; i}[{4\over\sqrt{5}}{\bar\epsilon}
{\gamma}_{ijk}\chi + {\bar\epsilon}{\tau}^{\underline7}{\gamma}^{l}{\gamma}_{ijk}{\lambda}_{l}]{\Pi}_{J}^{\; \; j}{\Pi}_{K}^{\;
\; k}F_{\mu\nu}^{JK}e^{-{\sigma\over\sqrt{10}}}
\nonumber\\
&& -{i\over8\sqrt{6}m}{\Pi}_{I}^{\; \;
i}[4{\bar\epsilon}{\gamma}_{ijk}
{\psi}_{\mu} +
{2\over\sqrt{5}}{\bar\epsilon}{\gamma}_{ijk}{\tau}^{\underline7}{\tau}_{\mu}
{\chi}+2{\bar\epsilon}{\tau}_{\mu}{\gamma}^{l}{\gamma}_{ijk}{\lambda}_{l}]
{\Pi}_{J}^{\; \; j}{\Pi}_{K}^{\; \; k}F_{\nu}^{JK}e^{3\sigma\over2\sqrt{10}}
\nonumber\\
&& -{i\sqrt{3}\over{12m}}{\delta}_{IJ}{\Pi}^{-1 \; \; J}_{\; \; \; \; i}
{\nabla}_{\mu}[{3\over\sqrt{5}}{\bar\epsilon}{\tau}_{\nu}{\gamma}^{i}{\chi}
e^{-{\sigma\over2\sqrt{10}}}+2{\bar\epsilon}{\tau}_{\nu}{\tau}^{\underline7}
{\lambda}^{i}e^{-{\sigma\over2\sqrt{10}}}
-2{\bar\epsilon}{\tau}^{\underline7}{\gamma}^{i}{\psi}_{\nu}
e^{-{\sigma\over2\sqrt{10}}}]
\nonumber\\
&& -{i\sqrt{3}\over{6}}{\delta}_{IJ}{\Pi}^{-1 \; \; J}_{\; \; \; \; i}Q^{[ij]}
[2{\bar\epsilon}{\tau}_{\mu}{\gamma}_{j}{\psi}_{\nu}
-{1\over\sqrt{5}}{\bar\epsilon}{\tau}_{\mu\nu}{\tau}^{\underline7}{\gamma}_{j}
{\chi}+{\bar\epsilon}{\tau}_{\mu\nu}{\lambda}_{j}]
e^{2\sigma\over\sqrt{10}}
\nonumber\\
&& -{1\over12\sqrt{10}m}{\bar\epsilon}{\tau}^{\rho}{\chi}
P_{\mu\nu\rho\alpha\beta\gamma IJ}^{-1}
\big [{\epsilon}^{\alpha\beta\gamma\delta\epsilon\eta}
{\tilde G}_{\delta\epsilon\eta}^{\; \; \; \; \; \;J} \big ]
e^{-{3\sigma\over2\sqrt{10}}}.
\eea

\newpage
\section{Appendix B}

In this section we present the explicit expression for the D=6
Chern-Simons term ${\Omega}(B)$\footnote{the trace is over the adjoint
representation of SO(5).}  
\bea 4{\Omega}(B) &=&
{\epsilon}^{\mu\nu\rho\sigma\lambda\tau}Tr \big \{
B_{\mu}F_{\nu\rho}F_{\sigma\lambda}F_{\tau}+B_{\mu}F_{\nu\rho}F_{\sigma}
F_{\lambda\tau}+B_{\mu}F_{\nu}F_{\rho\sigma}F_{\lambda\tau}
+{1\over2}B_{0}F_{\mu\nu}F_{\rho\sigma}F_{\lambda\tau} \nonumber\\ &&
-{4\over5}g {\lbrack} B_{\mu}B_{\nu}B_{\rho}F_{\sigma\lambda}F_{\tau}
+B_{\mu}B_{\nu}B_{\rho}F_{\sigma}F_{\lambda\tau}
+{1\over2}B_{\mu}B_{\nu}B_{0}F_{\rho\sigma}F_{\lambda\tau}
-{1\over2}B_{\mu}B_{0}B_{\nu}F_{\rho\sigma}F_{\lambda\tau} \nonumber\\
&& +{1\over2}B_{0}B_{\mu}B_{\nu}F_{\rho\sigma}F_{\lambda\tau}{\rbrack}
-{2\over5}g{\lbrack}B_{\mu}B_{\nu}F_{\rho\sigma}B_{\lambda}F_{\tau}
+{1\over2}B_{\mu}B_{\nu}F_{\rho\sigma}B_{0}F_{\lambda\tau}
-B_{\mu}B_{\nu}F_{\rho}B_{\sigma}F_{\lambda\tau} \nonumber\\ &&
-{1\over2}B_{\mu}B_{0}F_{\nu\rho}B_{\sigma}F_{\lambda\tau}
+{1\over2}B_{0}B_{\mu}F_{\nu\rho}B_{\sigma}F_{\lambda\tau}{\rbrack}
+{4\over5}{g^2}{\lbrack}B_{\mu}B_{\nu}B_{\rho}B_{\sigma}B_{\lambda}F_{\tau}
\nonumber\\ &&
+{1\over2}B_{\mu}B_{\nu}B_{\rho}B_{\sigma}B_{0}F_{\lambda\tau}
-{1\over2}B_{\mu}B_{\nu}B_{\rho}B_{0}B_{\sigma}F_{\lambda\tau}
+{1\over2}B_{\mu}B_{\nu}B_{0}B_{\rho}B_{\sigma}F_{\lambda\tau}
\nonumber\\ &&
-{1\over2}B_{\mu}B_{0}B_{\nu}B_{\rho}B_{\sigma}F_{\lambda\tau}
+{1\over2}B_{0}B_{\mu}B_{\nu}B_{\rho}B_{\sigma}F_{\lambda\tau}{\rbrack}
-{4\over5}{g^3}{\lbrack}B_{\mu}B_{\nu}B_{\rho}B_{\sigma}B_{\lambda}
B_{\tau}B_{0}{\rbrack} \big \} \nonumber\\ &&
-{1\over{2}}{\epsilon}^{\mu\nu\rho\sigma\lambda\tau}Tr(
B_{\mu}F_{\nu}+{1\over2}B_{0}F_{\mu\nu}-gB_{0}B_{\mu}B_{\nu})
Tr(F_{\rho\sigma}F_{\lambda\tau}) \nonumber\\ &&
-{1\over{2}}{\epsilon}^{\mu\nu\rho\sigma\lambda\tau}Tr(
B_{\mu}F_{\nu\rho}-{4\over3}gB_{\mu}B_{\nu}B_{\rho})
Tr(F_{\sigma\lambda}F_{\tau}).  \eea

\newpage
\section{Appendix C}

Here we present the complete Lagrangian and supersymmetry
transformation laws of D=6 maximal supergravity ({\it i.e.} the $g=0$
limit of the model of appendix A). The field
content is 1 graviton $g_{\mu\nu}$, 5 2-index antisymmetric tensor potentials
$C_{\mu\nu I}$, (10+5+1) vectors ($B_{\mu I}^{\; \; \; \; J}, S_{\mu
I}, A_{\mu}$),
(14+10+1) scalars (${\Pi}_{I}^{\; \; j}, B_{0I}^{\; \; \; \; J},
{\sigma}$), 4 gravitini ${\psi}_{\mu}$ and (16+4) spin 1/2 fermions
(${\lambda}_{i}, {\chi}$). ${\psi}_{\mu}$, ${\lambda}_{i}$ and ${\chi}$
are all D=6 USp(4) symplectic Majorana spinors and the spinor
inversion formula reads
\bea
{\bar\lambda}{\tau}^{{\alpha}_{1}\dots{\alpha}_{n}}
{\gamma}^{{i}_{1}\dots{i}_{r}}{\chi} = (-1)^{n(n+1)\over2}(-1)^{r(r-1)\over2}
{\bar\chi}{\tau}^{{\alpha}_{1}\dots{\alpha}_{n}}
{\gamma}^{{i}_{1}\dots{i}_{r}}{\lambda}
\nonumber
\eea
where ${\gamma}^{i}$ are the D=5 Dirac matrices. The Lagrangian of D=6
maximal supergravity, neglecting quartic fermion terms, is:

\bea
  e^{-1}{\cal L}_6 &=& R-{1\over4}e^{-{5\sigma\over\sqrt{10}}}(f_{\mu\nu})^2
  -{1\over{12}}e^{-{2\sigma\over\sqrt{10}}}({\Pi}^{-1 \; \; I}_{\; \; \;
  \; i}H_{\mu\nu\rho I})^2
  -{1\over4}e^{-{\sigma\over\sqrt{10}}}({\Pi}_{I}^{\; \; i}{\Pi}_{J}^{\;
  \; j} F_{\mu\nu}^{IJ})^2
\nonumber\\
&& -{1\over{4}}e^{3\sigma\over\sqrt{10}}({\Pi}^{-1 \; \; I}_{\; \; \;
  \; i}G_{\mu\nu I})^2
  -{1\over2}e^{4\sigma\over\sqrt{10}}({\Pi}_{I}^{\; \; i}{\Pi}_{J}^{\;
  \; j} F_{\mu}^{IJ})^2 
  -{1\over2}({\partial}_{\mu}{\sigma})^{2}-P_{\mu ij}P^{\mu ij}
\nonumber\\
&&-{{e^{-1}}\over36\sqrt{2}}{\epsilon}^{\mu\nu\rho\sigma\lambda\tau}
B_{0}^{\; \; IJ}H_{\mu\nu\rho I}H_{\sigma\lambda\tau J}
-{{e^{-1}}\over6\sqrt{2}}{\epsilon}^{\mu\nu\rho\sigma\lambda\tau}
H_{\mu\nu\rho I}B_{\sigma}^{\; \; IJ}G_{\lambda\tau J}
\nonumber\\
&&-{\bar\psi}_{\mu}{\tau}^{\mu\nu\rho}{\nabla}_{\nu}{\psi}_{\rho}
   -{\bar\chi}{\tau}^{\mu}{\nabla}_{\mu}{\chi}
   -{\bar\lambda}^{i}{\tau}^{\mu}{\nabla}_{\mu}{\lambda}_{i}
+{\bar\psi}_{\mu}{\tau}^{\nu}{\tau}^{\mu}{\gamma}^{i}{\lambda}^{j}P_{\nu ij}
\nonumber\\
&&+{1\over{4\sqrt{10}}}{\bar\psi}_{\mu}g^{\mu\nu}{\tau}^{\lambda}{\psi}_{\nu}{\partial}_{\lambda}\sigma
-{1\over{\sqrt{2}}}{\bar\psi}_{\mu}{\tau}^{\nu}{\tau}^{\mu}{\tau}^{\underline7}
{\chi}{\partial}_{\nu}\sigma
\nonumber\\
&&-{1\over{8}}{\bar\psi}_{\mu}[2{\tau}^{\mu\nu\rho}{\tau}^{\lambda}+
{\tau}^{\mu\rho}{\tau}^{\nu\lambda}]{\tau}^{\underline7}{\psi}_{\rho}f_{\nu\lambda}e^{-{5\sigma\over2\sqrt{10}}}
-{{\sqrt5}\over8}{\bar\psi}_{\mu}{\tau}^{\nu\lambda}{\tau}^{\mu}{\chi}f_{\nu\lambda}e^{-{5\sigma\over2\sqrt{10}}}
\nonumber\\
&&-{3\over{16}}{\bar\chi}{\tau}^{\underline7}{\tau}^{\nu\lambda}{\chi}f_{\nu\lambda}e^{-{5\sigma\over2\sqrt{10}}}
-{1\over{8}}{\bar\lambda}^{i}{\tau}^{\underline7}{\tau}^{\nu\lambda}{\lambda}_{i}f_{\nu\lambda}e^{-{5\sigma\over2\sqrt{10}}}
\nonumber\\
&&+{1\over{8\sqrt2}}{\bar\psi}_{\mu}[{\tau}^{\mu\nu\rho\sigma}-2{g}^{\mu\nu}
{g}^{\rho\sigma}]{\gamma}_{ij}{\psi}_{\sigma}{\Pi}_{I}^{\; \; i}{\Pi}_{J}^{\;
  \; j}F_{\nu\rho}^{IJ}e^{-{\sigma\over2\sqrt{10}}}
\nonumber\\
&& +{1\over{8\sqrt{10}}}{\bar\psi}_{\mu}[2{g}^{\mu\nu}{\tau}^{\rho}+{\tau}^{\mu\nu\rho}]{\tau}^{\underline7}{\gamma}_{ij}{\chi}{\Pi}_{I}^{\; \; i}{\Pi}_{J}^{\;
  \; j}F_{\nu\rho}^{IJ}e^{-{\sigma\over2\sqrt{10}}}
\nonumber\\
&& -{11\over{80\sqrt{2}}}{\bar\chi}{\tau}^{\nu\rho}{\gamma}_{ij}{\chi}{\Pi}_{I}^{\;
  \; i}{\Pi}_{J}^{\; \; j}F_{\nu\rho}^{IJ}e^{-{\sigma\over2\sqrt{10}}}
+{1\over{2\sqrt{2}}}{\bar\psi}_{\mu}{\tau}^{\nu\rho}{\tau}^{\mu}{\gamma}_{i}{\lambda}_{j}{\Pi}_{I}^{\; \; i}{\Pi}_{J}^{\; \; j}F_{\nu\rho}^{IJ}
e^{-{\sigma\over2\sqrt{10}}}
\nonumber\\
&& -{1\over{2\sqrt{10}}}{\bar\chi}{\tau}^{\nu\rho}{\tau}^{\underline7}{\gamma}_{i}{\lambda}_{j}{\Pi}_{I}^{\; \; i}{\Pi}_{J}^{\; \; j}F_{\nu\rho}^{IJ}
e^{-{\sigma\over2\sqrt{10}}}
+{1\over{16\sqrt{2}}}{\bar\lambda}_{i}{\gamma}^{j}{\gamma}_{kl}{\gamma}^{i}{\tau}^{\nu\rho}{\lambda}_{j}{\Pi}_{I}^{\; \; k}{\Pi}_{J}^{\; \; l}F_{\nu\rho}^{IJ}
e^{-{\sigma\over2\sqrt{10}}}
\nonumber\\
&&-{1\over{4\sqrt{2}}}{\bar\psi}_{\mu}{\tau}^{\mu\nu\rho}{\tau}^{\underline7}{\gamma}_{ij}{\psi}_{\rho}{\Pi}_{I}^{\; \; i}{\Pi}_{J}^{\; \; j}F_{\nu}^{IJ}
e^{2\sigma\over\sqrt{10}}
-{1\over{\sqrt{10}}}{\bar\psi}_{\mu}{\tau}^{\mu}{\tau}^{\nu}{\gamma}_{ij}{\chi}{\Pi}_{I}^{\; \; i}{\Pi}_{J}^{\; \; j}F_{\nu}^{IJ}e^{2\sigma\over\sqrt{10}}
\nonumber\\
&& +{7\over{20\sqrt{2}}}{\bar\chi}{\tau}^{\underline7}{\tau}^{\nu}{\gamma}_{ij}{\chi}{\Pi}_{I}^{\; \; i}{\Pi}_{J}^{\; \; j}F_{\nu}^{IJ}e^{2\sigma\over\sqrt{10}}
-{1\over{\sqrt{2}}}{\bar\psi}_{\mu}{\tau}^{\nu}{\tau}^{\mu}{\tau}^{\underline7}{\gamma}_{i}{\lambda}_{j}{\Pi}_{I}^{\;
  \; i}{\Pi}_{J}^{\; \; j}F_{\nu}^{IJ}
e^{2\sigma\over\sqrt{10}}
\nonumber\\
&& +{4\over{\sqrt{10}}}{\bar\chi}{\tau}^{\nu}{\gamma}_{i}{\lambda}_{j}{\Pi}_{I}^{\;
  \; i}{\Pi}_{J}^{\; \; j}F_{\nu}^{IJ}e^{2\sigma\over\sqrt{10}}
+{1\over{8\sqrt{2}}}{\bar\lambda}_{i}{\gamma}^{j}{\gamma}_{kl}{\gamma}^{i}{\tau}^{\nu}{\tau}^{\underline7}{\lambda}_{j}{\Pi}_{I}^{\;
  \; k}{\Pi}_{J}^{\; \; l}F_{\nu}^{IJ}e^{2\sigma\over\sqrt{10}}
\nonumber\\
&& +{i\over{8}}{\bar\psi}_{\mu}[{\tau}^{\mu\nu\rho\lambda}+
{g}^{\mu\nu}{g}^{\rho\lambda}]{\tau}^{\underline7}{\gamma}^{i}
{\psi}_{\lambda}{\Pi}^{-1 \; \; I}_{\; \; \; \; i}G_{\nu\rho I}
e^{3\sigma\over2\sqrt{10}}
\nonumber\\
&& +{3i\over{8\sqrt{5}}}{\bar\psi}_{\mu}[{\tau}^{\mu\nu\rho}-2
{g}^{\mu\nu}{\tau}^{\rho}]{\gamma}^{i}{\chi}{\Pi}^{-1 \; \; I}_{\;
  \; \; \; i}G_{\nu\rho I}e^{3\sigma\over2\sqrt{10}}
\nonumber\\
&& +{i\over{80}}{\bar\chi}{\tau}^{\underline7}{\tau}^{\nu\rho}
{\gamma}^{i}{\chi}{\Pi}^{-1 \; \; I}_{\; \; \; \; i}G_{\nu\rho I}
e^{3\sigma\over2\sqrt{10}}
+{i\over{4}}{\bar\psi}_{\mu}[{\tau}^{\mu\nu\rho}-2{g}^{\mu\nu}
{\tau}^{\rho}]{\tau}^{\underline7}{\lambda}^{i}{\Pi}^{-1 \; \; I}_{\; \; \; \; i}G_{\nu\rho I}e^{3\sigma\over2\sqrt{10}}
\nonumber\\
&& -{3i\over{4\sqrt{5}}}{\bar\chi}{\tau}^{\nu\rho}{\lambda}^{i}{\Pi}^{-1
  \; \; I}_{\; \; \; \; i}G_{\nu\rho I}e^{3\sigma\over2\sqrt{10}}
+{i\over{8}}{\bar\lambda}^{j}{\tau}^{\nu\rho}{\tau}^{\underline7}
{\gamma}^{i}{\lambda}_{j}{\Pi}^{-1 \; \; I}_{\; \; \; \; i}
G_{\nu\rho I}e^{3\sigma\over2\sqrt{10}}
\nonumber\\
&&+{i\over{24}}{\bar\psi}_{\sigma}[{\tau}^{\sigma\mu\nu\rho\lambda}+6{g}^{\sigma\mu}{\tau}^{\nu}{g}^{\rho\lambda}]{\gamma}^{i}{\psi}_{\lambda}
{\Pi}^{-1 \; \; I}_{\; \; \; \; i}H_{\mu\nu\rho I}e^{-{\sigma\over\sqrt{10}}}
\nonumber\\
&&+{i\over{12\sqrt{5}}}{\bar\psi}_{\sigma}[{\tau}^{\sigma\mu\nu\rho}-3{g}^{\sigma\mu}{\tau}^{\nu\rho}]{\tau}^{\underline7}{\gamma}^{i}{\chi}
{\Pi}^{-1 \; \; I}_{\; \; \; \; i}H_{\mu\nu\rho I}e^{-{\sigma\over\sqrt{10}}}
\nonumber\\
&&+{i\over{40}}{\bar\chi}{\tau}^{\mu\nu\rho}{\gamma}^{i}{\chi}
{\Pi}^{-1 \; \; I}_{\; \; \; \; i}H_{\mu\nu\rho I}e^{-{\sigma\over\sqrt{10}}}
-{i\over{12}}{\bar\psi}_{\sigma}[{\tau}^{\sigma\mu\nu\rho}-3{g}^{\sigma\mu}
{\tau}^{\nu\rho}]{\lambda}^{i}
{\Pi}^{-1 \; \; I}_{\; \; \; \; i}H_{\mu\nu\rho I}e^{-{\sigma\over\sqrt{10}}}
\nonumber\\
&&-{i\over6\sqrt{5}}{\bar\chi}{\tau}^{\underline7}{\tau}^{\mu\nu\rho}
{\lambda}^{i}{\Pi}^{-1 \; \; I}_{\; \; \; \; i}H_{\mu\nu\rho I}e^{-{\sigma\over\sqrt{10}}}
-{i\over{24}}{\bar\lambda}^{j}{\tau}^{\mu\nu\rho}{\gamma}^{i}{\lambda}_{j}
{\Pi}^{-1 \; \; I}_{\; \; \; \; i}H_{\mu\nu\rho I}e^{-{\sigma\over\sqrt{10}}}.
\eea
where 
\bea
&  f_{2}=dA \; \; \; \; \; \; \; \; G_{2I}=dS_{1I}
\nonumber\\
&  F_{2 I}^{\; \; \; J}= dB_{1 I}^{\; \; \; J}+B_{0 I}^{\; \; \;
   J}dA \; \; \; \; \; \; \; F_{1 I}^{\; \; \; J}= dB_{0 I}^{\; \; \; J} 
\nonumber\\
&  {H}_{\mu\nu\rho I} = 3({\partial}_{[ \mu}C_{\nu\rho ]I}
  +{1\over2}{G}_{[ \mu\nu I}{A}_{\rho ]})
\nonumber\\
&  P_{\mu ij}={\Pi}^{-1 \; I}_{\; \; \; \; (i}
  {\partial}_{\mu}^{\ }{\Pi}_{Ij)}^{\ }
\nonumber\\
&  Q_{\mu ij}={\Pi}^{-1 \; I}_{\; \; \; \; [i}
  {\partial}_{\mu}^{\ }{\Pi}_{Ij]}^{\ },
\eea
and
\bea
{\nabla}_{\mu}{\xi}_{\nu i} = {\partial}_{\mu}{\xi}_{\nu i}+{1\over4}
{\omega}_{\mu\underline{\nu}\underline{\rho}}
{\tau}^{\underline{\nu}\underline{\rho}}
{\xi}_{\nu i}+{\Gamma}_{\mu\nu}^{\rho}{\xi}_{\rho i}+
{1\over4}{Q}_{\mu jk}{\gamma}^{jk}{\xi}_{\nu i}+
{Q}_{\mu i}^{\; \; \; j}{\xi}_{\nu j},
\eea
for a general vector spinor, ${\xi}_{\nu i}$, of both SO(5,1) and SO(5)$_{c}$.
${\psi}_{\mu}$ and ${\chi}$ are SO(5)$_{c}$ spinors and
  ${\lambda}_{i}$ is an SO(5)$_{c}$ vector spinor.
The supersymmetry transformations of the fields neglecting terms cubic
in fermion fields are:

\bea 
{\delta}{\psi}_{\mu} &=& {\nabla}_{\mu}{\epsilon}
-{1\over32}[{\tau}_{\mu}^{\; \; \nu\rho}-6{\delta}_{\mu}^{\nu}{\tau}^{\rho}]
{\tau}^{\underline7}{\epsilon}f_{\nu\rho}e^{-{5\sigma\over{2\sqrt{10}}}}
-{1\over{32\sqrt{2}}}[{\tau}_{\mu}^{\; \; \nu\rho}-6{\delta}_{\mu}^{\nu}{\tau}^{\rho}]{\gamma}_{ij}{\epsilon}{\Pi}_{I}^{\;
  \; i}{\Pi}_{J}^{\; \; j}F_{\nu\rho}^{IJ}
e^{-{\sigma\over{2\sqrt{10}}}}
\nonumber\\
&&+ {1\over{4\sqrt{2}}}{\tau}^{\underline7}{\gamma}_{ij}{\epsilon}{\Pi}_{I}^{\;
  \; i}{\Pi}_{J}^{\; \; j}F_{\mu}^{IJ}
e^{2\sigma\over{\sqrt{10}}}
+ {1\over{2\sqrt{10}}}{\epsilon}{\partial}_{\mu}{\sigma}
-{i\over{32}}[{\tau}_{\mu}^{\; \; \nu\rho}-6{\delta}_{\mu}^{\nu}
{\tau}^{\rho}]{\tau}^{\underline7}{\gamma}^{i}{\epsilon}{\Pi}^{-1 \;
  \; I}_{\; \; \; \; i}G_{\nu\rho I}e^{3\sigma\over2\sqrt{10}}
\nonumber\\
&&+{i\over{48}}[{\tau}_{\mu}^{\; \; \nu\rho\sigma}-3{\delta}_{\mu}^{\nu}
{\tau}^{\rho\sigma}]{\gamma}^{i}{\epsilon}{\Pi}^{-1 \; \; I}_{\;
  \; \; \; i}H_{\nu\rho\sigma I}
e^{-{\sigma\over\sqrt{10}}},
\eea
\bea
{\delta}{\chi} &=& -{\sqrt{5}\over{16}}{\tau}^{\mu\nu}{\epsilon}f_{\mu\nu}
e^{-{5\sigma\over{2\sqrt{10}}}}
+{1\over{2\sqrt{2}}}{\tau}^{\mu}{\tau}^{\underline7}{\epsilon}
{\partial}_{\mu}{\sigma}
-{1\over{16\sqrt{10}}}{\tau}^{\mu\nu}{\tau}^{\underline7}{\gamma}_{ij}
{\epsilon}{\Pi}_{I}^{\; \; i}{\Pi}_{J}^{\; \; j}F_{\mu\nu}^{IJ}
e^{-{\sigma\over{2\sqrt{10}}}}
\nonumber\\
&&-{1\over{2\sqrt{10}}}{\tau}^{\mu}{\gamma}_{ij}{\epsilon}
{\Pi}_{I}^{\; \; i}{\Pi}_{J}^{\; \; j}F_{\mu}^{IJ}
e^{2\sigma\over{\sqrt{10}}}
+{3i\over{16\sqrt{5}}}{\tau}^{\mu\nu}{\gamma}^{i}{\epsilon}
{\Pi}^{-1 \; \; I}_{\; \; \; \; i}G_{\mu\nu I}e^{3\sigma\over2\sqrt{10}}
\nonumber\\
&&-{i\over{24\sqrt{5}}}{\tau}^{\mu\nu\rho}{\tau}^{\underline7}{\gamma}^{i}
{\epsilon}{\Pi}^{-1 \; \; I}_{\; \; \; \; i}H_{\mu\nu\rho I}
e^{-{\sigma\over\sqrt{10}}},
\eea
\bea
{\delta}{\lambda}_{i} &=&
{1\over{16\sqrt{2}}}{\tau}^{\mu\nu}[{\gamma}_{kl}{\gamma}_{i}-{1\over5}
{\gamma}_{i}{\gamma}_{kl}]{\epsilon}{\Pi}_{I}^{\; \; k}{\Pi}_{J}^{\;
\; l}F_{\mu\nu}^{IJ}e^{-{\sigma\over{2\sqrt{10}}}}
+{1\over2}{\tau}^{\mu}{\gamma}^{j}{\epsilon}P_{\mu ij}
\nonumber\\
&&+{1\over{8\sqrt{2}}}{\tau}^{\mu}{\tau}^{\underline7}
[{\gamma}_{kl}{\gamma}_{i}-{1\over5}{\gamma}_{i}{\gamma}_{kl}]{\epsilon}
{\Pi}_{I}^{\; \; k}{\Pi}_{J}^{\; \; l}F_{\mu}^{IJ}e^{2\sigma\over{\sqrt{10}}}
\nonumber\\
&&-{i\over{40}}{\tau}^{\mu\nu}{\tau}^{\underline7}[{\gamma}_{i}^{\;
j}-4{\delta}_{i}^{\; j}]{\epsilon}{\Pi}^{-1 \; \; I}_{\; \; \; \;
j}G_{\mu\nu I}e^{3\sigma\over2\sqrt{10}}
\nonumber\\
&&+{i\over{120}}{\tau}^{\mu\nu\rho}[{\gamma}_{i}^{\;
j}-4{\delta}_{i}^{\; j}]{\epsilon}{\Pi}^{-1 \; \; I}_{\; \; \; \; j}
H_{\mu\nu\rho I}e^{-{\sigma\over\sqrt{10}}},
\eea
\bea
{\delta\sigma} = -{1\over\sqrt2}{\bar\epsilon}{\tau}^{\underline7}\chi
e^{\sigma\over\sqrt{10}}, \; \; \; \; \; \; \; \; \; \; \; \; \; 
{\Pi}^{-1 \; \; I}_{\; \; \; \; i}{\delta}
{\Pi}_{I}^{\; \; j}= {1\over4}[{\bar\epsilon}{\gamma}_{i}{\lambda}^{j}+
{\bar\epsilon}{\gamma}^{j}{\lambda}_{i}]e^{\sigma\over\sqrt{10}},
\eea
\bea
{\delta}A_{\mu} = {1\over2}{\bar\epsilon}{\tau}^{\underline7}{\psi}_{\mu}
e^{7\sigma\over2\sqrt{10}}+{1\over4\sqrt5}{\bar\epsilon}{\tau}_{\mu}\chi
e^{7\sigma\over2\sqrt{10}},
\eea
\bea
{\delta}{e^{\underline{\mu}}}^{\ }_{\mu}={1\over2}{\bar\epsilon}
{\tau}^{\underline{\mu}}{\psi}_{\mu}e^{\sigma\over\sqrt{10}}
+ {1\over\sqrt{5}}{\bar\epsilon}{\tau}^{\underline{\mu}}{\chi}A_{\mu}
e^{-{3\sigma\over2\sqrt{10}}}
+{1\over4\sqrt{5}}{\bar\epsilon}{\tau}_{\mu}^{\; \; \nu}
{e^{\underline{\mu}}}^{\ }_{\nu}{\tau}^{\underline7}{\chi}
e^{\sigma\over\sqrt{10}},
\eea
\bea
{\Pi}_{I}^{\; \; i}{\Pi}_{J}^{\; \; j}{\delta}B_{0}^{\; \; IJ} &=&
{1\over\sqrt{10}}{\bar\epsilon}{\gamma}^{ij}{\chi}e^{-{\sigma\over\sqrt{10}}}
-{1\over\sqrt{5}}{\bar\epsilon}{\tau}^{\mu}\chi{\Pi}_{I}^{\; \;
i}{\Pi}_{J}^{\; \; j}B_{\mu}^{\; \; IJ}e^{-{3\sigma\over2\sqrt{10}}} 
\nonumber\\
&&+{1\over4\sqrt{2}}{\bar\epsilon}{\tau}^{\underline7}{\gamma}^{k}{\gamma}^{ij}
{\lambda}_{k}e^{-{\sigma\over\sqrt{10}}},
\eea
\bea
{\Pi}_{I}^{\; \; i}{\Pi}_{J}^{\; \; j}{\delta}B_{\mu}^{\; \; IJ} &=&
{1\over2\sqrt{2}}{\bar\epsilon}{\gamma}^{ij}{\psi}_{\mu}
e^{3\sigma\over2\sqrt{10}}
+{1\over4\sqrt{10}}{\bar\epsilon}{\gamma}^{ij}{\tau}^{\underline7}{\tau}_{\mu}
{\chi}e^{3\sigma\over2\sqrt{10}}
+{1\over4\sqrt{2}}{\bar\epsilon}{\tau}_{\mu}{\gamma}^{k}{\gamma}^{ij}
{\lambda}_{k}e^{3\sigma\over2\sqrt{10}}
\nonumber\\
&&-{1\over2}{\bar\epsilon}{\tau}^{\underline7}{\psi}_{\mu}{\Pi}_{I}^{\; \; i}{\Pi}_{J}^{\; \; j}B_{0}^{\; \; IJ}e^{7\sigma\over{2\sqrt{10}}}
-{1\over4\sqrt{5}}{\bar\epsilon}{\tau}_{\mu}\chi{\Pi}_{I}^{\; \; i}{\Pi}_{J}^{\; \; j}B_{0}^{\; \; IJ}e^{7\sigma\over{2\sqrt{10}}}
\nonumber\\
&&+{1\over\sqrt{5}}{\bar\epsilon}{\tau}^{\nu}A_{\mu}\chi{\Pi}_{I}^{\; \; i}{\Pi}_{J}^{\; \; j}B_{\nu}^{\; \; IJ}e^{-{3\sigma\over{2\sqrt{10}}}},
\eea
\bea
{\delta}G_{\mu\nu I} &=& -{2\over\sqrt{5}}{\bar\epsilon}{\tau}^{\rho}{\chi}
A_{\mu}G_{\nu\rho I}e^{-{3\sigma\over{2\sqrt{10}}}}
+{i\over4\sqrt{2}}{\Pi}_{I}^{\; \; i}[{4\over\sqrt{5}}{\bar\epsilon}
{\gamma}_{ijk}\chi + {\bar\epsilon}{\tau}^{\underline7}{\gamma}^{l}{\gamma}_{ijk}{\lambda}_{l}]{\Pi}_{J}^{\; \; j}{\Pi}_{K}^{\;
\; k}F_{\mu\nu}^{JK}e^{-{\sigma\over\sqrt{10}}}
\nonumber\\
&&+{i\over4\sqrt{2}}{\Pi}_{I}^{\; \; i}[4{\bar\epsilon}{\gamma}_{ijk}
{\psi}_{\mu} +
{2\over\sqrt{5}}{\bar\epsilon}{\gamma}_{ijk}{\tau}^{\underline7}{\tau}_{\mu}
{\chi}+2{\bar\epsilon}{\tau}_{\mu}{\gamma}^{l}{\gamma}_{ijk}{\lambda}_{l}]
{\Pi}_{J}^{\; \; j}{\Pi}_{K}^{\; \; k}F_{\nu}^{JK}e^{3\sigma\over2\sqrt{10}}
\nonumber\\
&&+{1\over\sqrt{5}}{\bar\epsilon}{\tau}^{\rho}{\chi}
H_{\mu\nu\rho I}e^{-{3\sigma\over2\sqrt{10}}},
\eea
\bea
{\delta}H_{\mu\nu\rho I} &=& {3\over2}{\bar\epsilon}{\tau}^{\underline7}
{\psi}_{\mu}G_{\nu\rho I}e^{7\sigma\over{2\sqrt{10}}}
+{3\over4\sqrt{5}}{\bar\epsilon}{\tau}_{\mu}{\chi}G_{\nu\rho I}
e^{7\sigma\over{2\sqrt{10}}}
+{3\over\sqrt{5}}{\bar\epsilon}{\tau}^{\lambda}{\chi}
A_{\mu}H_{\nu\rho\lambda I}e^{-{3\sigma\over{2\sqrt{10}}}}
\nonumber\\
&&-{3i\over4\sqrt{2}}{\Pi}_{I}^{\; \; i}[2{\bar\epsilon}{\gamma}_{ijk}
{\psi}_{\mu}
+{1\over\sqrt{5}}{\bar\epsilon}{\gamma}_{ijk}{\tau}^{\underline7}{\tau}_{\mu}
{\chi}+{\bar\epsilon}{\tau}_{\mu}{\gamma}^{l}{\gamma}_{ijk}{\lambda}_{l}]
{\Pi}_{J}^{\; \; j}{\Pi}_{K}^{\; \; k}F_{\nu\rho}^{JK}
e^{3\sigma\over2\sqrt{10}}.
\eea

\section{Appendix D}

The bosonic sector of the D=5 N=4 SU(2) gauged supergravity was
obtained in \cite{PMC} by reduction on T$^2$ of D=7 SU(2) gauged supergravity
\cite{TvN}. The bosonic sector of the Lagrangian is \cite{PMC}:
\bea
    e^{-1}{\cal L} &=& R_{5} -{1\over2}{|d\psi|}^{2}
             -{1\over4}e^{-{2\psi\over\sqrt{6}}} [ |F_{2}|^{2} +
             |G_{2}^{(1)}|^{2} + |G_{2}^{(2)}|^{2} ]
             -{1\over4}e^{4\psi\over\sqrt{6}} |C_{2}|^{2} +
             4{\alpha}^{2} e^{2\psi\over\sqrt{6}} 
\nonumber\\
&&             -{{e^{-1}}\over8}
             \epsilon^{\mu\nu\rho\sigma\tau} C_{\mu} [
             G_{\nu\rho}^{(1)} G_{\sigma\tau}^{(1)} +
             G_{\nu\rho}^{(2)} G_{\sigma\tau}^{(2)} +
             Tr(F_{\nu\rho}F_{\sigma\tau}) ] 
\label{pmclag1}
\eea          
where $G_{2}^{(p)} = dB_{1}^{(p)}$, ($p$=1,2), $C_2=dC_1$ and
${F_{\mu\nu i}}^{j} = 2(\partial_{[\mu}{A_{\nu ] i}}^{j} +
i\alpha{A_{[\mu |i|}}^{k}{A_{\nu ] k}}^{j})$, ($i$,$j$=1,2). Hence the
global SO(2) symmetry is manifest. The SU(2)$\times$U(1) gauged model
of Romans \cite{romans2} contains a pair of 2-index potentials
$B_{\mu\nu}^{\alpha}$ instead of the vectors $B_{\mu}^{(p)}$ in
(\ref{pmclag1}). This allows the U(1) symmetry to be gauged. The bosonic
sector of Romans' model is the same as (\ref{pmclag1}) except there is an extra
term in the scalar potential proportional to the U(1) coupling
constant $g_1$ and instead of the terms

\be
        -{e\over4}{\xi}^{2}{|G_{\mu\nu}^{(1)}|}^{2}-{e\over4}{\xi}^{2}{|G_{\mu\nu}^{(2)}|}^{2}-{1\over8}{\epsilon}^{\mu\nu\rho\sigma\tau}C_{\mu}[G_{\nu\rho}^{(1)}G_{\sigma\tau}^{(1)}+G_{\nu\rho}^{(2)}G_{\sigma\tau}^{(2)}],
\label{pmclag2}
\ee
\noindent
where $\xi=e^{-{\psi\over\sqrt6}}$, Romans' model contains the terms
\be
 {\cal L}={\epsilon}^{\mu\nu\rho\sigma\tau}[{1\over{g_1}}{\epsilon}_{\alpha\beta}{B_{\mu\nu}^{\alpha}}D_{\rho}{B_{\sigma\tau}^{\beta}}]-e{\xi}^{2}B^{\mu\nu\alpha}B_{\mu\nu\alpha},
\label{romlag}
\ee
\noindent
where $\alpha$=1,2 and
$D_{\mu}{B_{\nu\rho}^{\alpha}}={\partial}_{\mu}{B_{\nu\rho}^{\alpha}}+{{g_1}\over2}{\epsilon}^{\alpha\beta}C_{\mu}{B_{\nu\rho\beta}}$.
We now show that the U(1) coupling constant $g_1$ {\it can} be taken
to zero\footnote{contrary to the statement that $g_1$ cannot be taken to
zero in \cite{romans2}.} and the Lagrangian (\ref{romlag}) in this limit 
becomes (\ref{pmclag2}), thus
showing the SU(2) gauged model (\ref{pmclag1}) is indeed the limit of Romans'
SU(2)$\times$U(1) gauged supergravity in which the U(1) gauging is
turned off.

Variation of (\ref{romlag}) w.r.t. $B_{\mu\nu}^{1}$ yields the field equation
\be
        e{\xi}^{2}B^{\mu\nu
        1}={1\over{g_1}}{\epsilon}^{\mu\nu\rho\sigma\tau}[{\partial}_{\rho}{B_{\sigma\tau}^{2}}-{{g_1}\over2}C_{\rho}{B_{\sigma\tau}^{1}}].
\ee
\noindent
Hence solving for ${B_{\mu\nu}^{1}}$ we have
\be
        P^{\mu\nu\sigma\tau}B_{\sigma\tau}^{1}={1\over{3g_{1}}}{\epsilon}^{\mu\nu\rho\sigma\tau}H_{\rho\sigma\tau},
\label{Beqn}
\ee
\noindent
where $H_{\mu\nu\rho}=3{\partial}_{\mu}B_{\nu\rho}^{2}$ and we have
defined the operator $P$ s.t.
\be
         P^{\mu\nu\rho\sigma}=e{\xi}^{2}g^{\mu\rho}g^{\nu\sigma}+{1\over2}{\epsilon}^{\mu\nu\rho\sigma\tau}C_{\tau}.
\label{opP}
\ee
\noindent
Hence we see $B_{\mu\nu}^{1}$ is an auxiliary field which we can
eliminate. Defining the inverse of $P$ s.t.
\be
        P^{\mu\nu\rho\sigma}(P^{-1})_{\rho\sigma\alpha\beta}={\delta}_{\alpha\beta}^{\mu\nu},
\ee
\noindent
we can solve (\ref{Beqn}) for $B_{\mu\nu}^{1}$ and substitute back in the
Lagrangian (\ref{romlag}). Hence we obtain a Lagrangian involving just the field
$B_{\mu\nu}^{2}$:
\be
      {\cal L}={1\over{9g_1}}[{\epsilon}^{\mu\nu\rho\sigma\tau}H_{\rho\sigma\tau}](P^{-1})_{\mu\nu\alpha\beta}[{\epsilon}^{\alpha\beta\gamma\delta\epsilon}H_{\gamma\delta\epsilon}]-B_{\mu\nu}^{2}P^{\mu\nu\sigma\tau}B_{\sigma\tau}^{2}.
\ee

Before we take $g_1$ to zero we must regurgitate a vector from $B_{\mu\nu}^{2}$
thus
\be
         B_{\mu\nu}^{2}\longrightarrow
         B_{\mu\nu}^{2}+{1\over{g_1}}G_{\mu\nu}^{(2)},
\ee
\noindent
where $G_{\mu\nu}^{(2)}=2{\partial}_{\mu}B_{\nu}^{(2)}$. The
Lagrangian therefore becomes:
\bea
  {\cal L} &=&-{1\over{{g_1}^{2}}}G_{\mu\nu}^{(2)}P^{\mu\nu\rho\sigma}G_{\rho\sigma}^{(2)}-{2\over{g_1}}G_{\mu\nu}^{(2)}P^{\mu\nu\rho\sigma}B_{\rho\sigma}^{2}-B_{\mu\nu}^{2}P^{\mu\nu\rho\sigma}B_{\rho\sigma}^{2}
\nonumber\\
&&+{1\over{9g_1}}[{\epsilon}^{\mu\nu\rho\sigma\tau}H_{\rho\sigma\tau}](P^{-1})_{\mu\nu\alpha\beta}[{\epsilon}^{\alpha\beta\gamma\delta\epsilon}H_{\gamma\delta\epsilon}].
\eea
\noindent
Now we see that after making the following field rescalings
$B_{\mu\nu}^{2}\longrightarrow {\sqrt{g_1}}B_{\mu\nu}^{2}$,
$G_{\mu\nu}^{(2)}\longrightarrow {g_1}G_{\mu\nu}^{(2)}$, the $g_1$=0 limit
can be obtained:
\be
{\cal L}=-G_{\mu\nu}^{(2)}P^{\mu\nu\rho\sigma}G_{\rho\sigma}^{(2)}+{1\over{9}}[{\epsilon}^{\mu\nu\rho\sigma\tau}H_{\rho\sigma\tau}](P^{-1})_{\mu\nu\alpha\beta}[{\epsilon}^{\alpha\beta\gamma\delta\epsilon}H_{\gamma\delta\epsilon}].
\ee
\noindent
The extra term in the scalar potential proportional to $g_1$ simply
vanishes in this limit.

In order to compare with the model (\ref{pmclag2}) (which contains 2 massless
vectors) we must dualise $H_{\gamma\delta\epsilon}$ to
$G_{\mu\nu}^{(1)}$. This is achieved by replacing
$3{\partial}_{\mu}B_{\nu\rho}^{2}$ by an independent field
$a_{\mu\nu\rho}$ and adding to ${\cal L}$ the term
\be
         \Delta {\cal L}=\kappa
{\epsilon}^{\mu\nu\rho\sigma\tau}a_{\mu\nu\rho}G_{\sigma\tau}^{(1)},
\ee
\noindent
where $\kappa$ is a constant and
$G_{\sigma\tau}^{(1)}=2{\partial}_{\sigma}B_{\tau}^{(1)}$. Variation of
${\cal L}+{\Delta}{\cal L}$ w.r.t. $a_{\mu\nu\rho}$ then gives
\be
         {2\over9}{(P^{-1})}_{\alpha\beta\mu\nu}[{\epsilon}^{\mu\nu\rho\sigma\tau}a_{\rho\sigma\tau}]= -\kappa G_{\alpha\beta}^{(1)}.
\ee
\noindent
Substituting back in ${\cal L}+{\Delta}{\cal L}$ (choosing $\kappa={1\over3}$ and
rescaling $G_{2}^{(2)}$) gives:
\be
         {\cal L}+{\Delta}{\cal L}=-{1\over4}G_{\mu\nu}^{(1)}P^{\mu\nu\rho\sigma}G_{\rho\sigma}^{(1)}-{1\over4}G_{\mu\nu}^{(2)}P^{\mu\nu\rho\sigma}G_{\rho\sigma}^{(2)},
\ee
\noindent
thus using the expression for $P^{\mu\nu\rho\sigma}$, (\ref{opP}), the
Lagrangian becomes identical to (\ref{pmclag2}). Hence we have shown that the
limit of the bosonic sector of D=5 N=4 SU(2)$\times$U(1) gauged
supergravity, in which the U(1) coupling constant is turned off, is
the D=5 N=4 SU(2) gauged model (\ref{pmclag1}) obtained in \cite{PMC}.

\end{document}